\pdfoutput=1

\documentclass[11pt]{article}
\RequirePackage[OT1]{fontenc}
\RequirePackage{amsthm,amsmath}
\RequirePackage[numbers]{natbib}
\RequirePackage[colorlinks,citecolor=blue,urlcolor=blue]{hyperref}

\usepackage[utf8]{inputenc} 
\usepackage[T1]{fontenc}    
\usepackage[colorlinks]{hyperref} 
\usepackage{url}            
\usepackage{booktabs}       
\usepackage{amsfonts}       
\usepackage{amsmath}
\usepackage{nicefrac}       
\usepackage{microtype}      
\usepackage{color}
\usepackage{bbm}
\usepackage{amssymb, enumerate}
\usepackage{makecell}
\usepackage{wrapfig}
\usepackage{extarrows}
\usepackage{arydshln}

\usepackage{caption}
\usepackage{multirow}

\usepackage{amsmath,amsfonts,amsthm, amssymb, enumerate}
\usepackage{lipsum}
\usepackage{caption}
\usepackage{amssymb,graphicx,subfigure}
\usepackage[bottom]{footmisc}
\usepackage[dvipsnames]{xcolor}
\usepackage{array,multirow}
\usepackage{enumitem}

\usepackage{amsmath,amssymb,amsfonts}
\usepackage{graphicx}
\usepackage{textcomp}

\usepackage{booktabs}
\usepackage{color}       
\usepackage{graphicx}    
\usepackage{subfigure,caption,multirow}

\usepackage{kotex}

\newcommand{\argmin}{\operatornamewithlimits{argmin}}

\newcommand{\supp}{\operatornamewithlimits{supp}}

\newcommand{\seg}{\operatornamewithlimits{SEG}}
\newcommand{\reg}{\operatornamewithlimits{REG}}
\newcommand{\bsa}{\operatornamewithlimits{BA}} 

\usepackage{algorithm,algpseudocode}
\algnewcommand\SEG{\item[{\textbf{1: Segmentation stage}}]}
\algnewcommand\REG{\item[{\textbf{2: Registration stage}}]}
\algnewcommand\AUG{\item[{\textbf{3: Augmentation stage}}]}  
\algnewcommand\Input{\item[$\textbf{Input:}$]}
\algnewcommand\Output{\item[{$\textbf{Output:}$}]}
\algnewcommand\Initialize{\item[{$\textbf{Initialize:}$}]}
\algnewcommand{\return}[1]{
	\State \textbf{return:}
	\Statex \hspace*{\algorithmicindent}\parbox[t]{.8\linewidth}{\raggedright #1}
}
 
\newcommand{\algrule}[1][.2pt]{\par\vskip.5\baselineskip\hrule height #1\par\vskip.5\baselineskip}      

\usepackage{amsmath}
\usepackage{commath}
\usepackage{hyperref}     

\usepackage{commath}
\usepackage{hyperref}     
\usepackage{textcomp}
\usepackage{xcolor}
\usepackage{url}
\usepackage{xr}
\usepackage{wrapfig,lipsum}
\usepackage{hyperref}    
\usepackage{amsthm, amssymb, enumerate}
\usepackage{caption}
\usepackage{enumitem}
\usepackage{kotex}
\usepackage{textcomp}
\usepackage{siunitx} 
\usepackage{wrapfig}
\usepackage{array}
\usepackage{textcomp}
\usepackage{stfloats}
\usepackage{url}
\usepackage{verbatim}
\usepackage{graphicx}
\usepackage{textcomp}
\usepackage{siunitx} 
\usepackage{stackengine}
\usepackage{setspace}
\usepackage[left=1.8cm,right=1.8cm,top=3.8cm,bottom=3.8cm,a4paper]{geometry}

\begin{document}
	\title{Enhancing Generative Networks for Chest Anomaly Localization through Automatic Registration-Based Unpaired-to-Pseudo-Paired Training Data Translation}
	\date{}
	\author{\parbox{\linewidth}{\centering
		Kyungsu Kim$^{1,2}$\thanks{Equal contribution}\,\,\thanks{Corresponding author: Kyungsu Kim (kskim.doc@gmail.com) and Myung Jin Chung (mjchung@skku.edu)}, Seong Je Oh$^{3}$\footnotemark[1], Chae Yeon Lim$^{5}$, Ju Hwan Lee$^{3}$, Tae Uk Kim$^{4}$, Myung Jin Chung$^{1,2,4,6}$\footnotemark[2]\\ 
		{\small $^{1}$Medical AI Research Center, Research Institute for Future Medicine, Samsung Medical Center, Seoul, Korea}\\
		{\small $^{2}$Department of Data Convergence and Future Medicine, Sungkyunkwan University School of Medicine, Seoul, Korea}\\
		{\small $^{3}$Department of Health Sciences and Technology, SAIHST, Sungkyunkwan University, Seoul, Korea}\\
            {\small $^{4}$Department of Digital Health, SAIHST, Sungkyunkwan University, Seoul, Korea}\\
            {\small $^{5}$Department of Medical Device Management and Research, SAIHST, Sungkyunkwan University, Seoul, Korea}\\
		{\small $^{6}$Department of Radiology, Samsung Medical Center, Sungkyunkwan University School of Medicine, Seoul, Korea}}
	} 
	\maketitle 
	
\begin{abstract}
Image translation based on a generative adversarial network (GAN-IT) is a promising method for the precise localization of abnormal regions in chest X-ray images (AL-CXR) even without the pixel-level annotation. However, heterogeneous unpaired datasets undermine existing methods to extract key features and distinguish normal from abnormal cases, resulting in inaccurate and unstable AL-CXR. To address this problem, we propose an improved two-stage GAN-IT involving registration and data augmentation. For the first stage, we introduce an advanced deep-learning-based registration technique that virtually and reasonably converts unpaired data into paired data for learning registration maps, by sequentially utilizing linear-based global and uniform coordinate transformation and AI-based non-linear coordinate fine-tuning. This approach enables independent and complex coordinate transformation of each detailed location of the lung while recognizing the entire lung structure, thereby achieving higher registration performance with resolving inherent artifacts caused by unpaired conditions. For the second stage, we apply data augmentation to diversify anomaly locations by swapping the left and right lung regions on the uniform registered frames, further improving the performance by alleviating imbalance in data distribution showing left and right lung lesions. The proposed method is model agnostic and shows consistent AL-CXR performance improvement in representative AI models. Therefore, we believe GAN-IT for AL-CXR can be clinically implemented by using our basis framework, even if learning data are scarce or difficult for the pixel-level disease annotation.
\end{abstract}

 \section{Introduction}
\label{sec:introduction}

Chest X-ray (CXR) imaging is used as a first-line test for identifying lung anomalies because it provides fast image generation with a low radiation dose \citep{self2013high}. However, it is difficult to accurately diagnose diseases under fine shading conditions. Existing systems and methods should be further improved as precise anomaly localization in CXR images (AL-CXR) are clinically important.
The recent application of deep learning (DL) to CXR diagnosis has substantially improved the detection of anomalies in patients  \citep{jain2021deep,nayak2021application,hammoudi2021deep}. 
Nevertheless, apart from simple discrimination, this approach does not guarantee the precise identification of subregions with anomalies. Although class activation maps and their variants have been typically used to achieve AL-CXR, they have limited resolution in blurry maps and show clinical limitations regarding interpretation (e.g., it is difficult to obtain normal class activation in a disease-free region). 

Generative adversarial network (GAN)-based image translation (GAN-IT) has emerged as a powerful technique for generating high-resolution images through adversarial learning \cite{motamed2021randgan, bhatt2021unsupervised, nakao2021unsupervised}. GAN-IT can translate between two domains even with unpaired data, offering significant advantages in scenarios where paired datasets are unavailable. This makes GAN-IT particularly effective in medical domains where obtaining paired datasets is challenging, and it is especially useful for various medical imaging tasks (e.g., anomaly detection, data augmentation, and modality translation). Therefore, GAN-IT can transform unmatched datasets (i.e., normal/abnormal image pairs taken in different environments from different patients) and convert actual CXR images containing diseased areas into normal CXR images, analyzing the differences to identify abnormal areas in the lungs \cite{siddiquee2019learning, baur2020steganomaly, wolleb2020descargan}. However, using unpaired datasets in GAN-IT does not ensure structural consistency, leading to significant discrepancies not only in diseased areas but also in non-diseased areas (e.g., ribs) between images \cite{yang2020unsupervised}. This fundamental problem arises from the inherent randomness in the one-to-one mapping learning process of GAN-IT, which results in the loss of critical structural information (i.e., anatomical features) between input and output images \cite{lu2019guiding}. The extreme differences between the datasets used for GAN-IT training for AL-CXR hinder the translation to optimal images (i.e., normal images without lung disease). Consequently, GAN-IT models face limitations in providing reliable translations when applied to unpaired datasets typically used in medical imaging, where structural information is crucial \cite{yang2020mri}.
 
To address these challenges, a promising approach recently proposed involves adding registration networks to GAN-IT learning. The addition of registration networks can improve some of the data consistency issues by converting unpaired data into a pseudo-pair format \cite{yang2020mri, kong2021breaking, arar2020unsupervised, chen2022unsupervised}. This approach aligns the structural elements of the image so that important anatomical features of the image are maintained during translation. These benefits enhance the alignment between input and output images, improving the overall accuracy and reliability of the GAN-IT model.

\begin{figure}[hbt!]
 \centering
\includegraphics[width=0.7\textwidth]{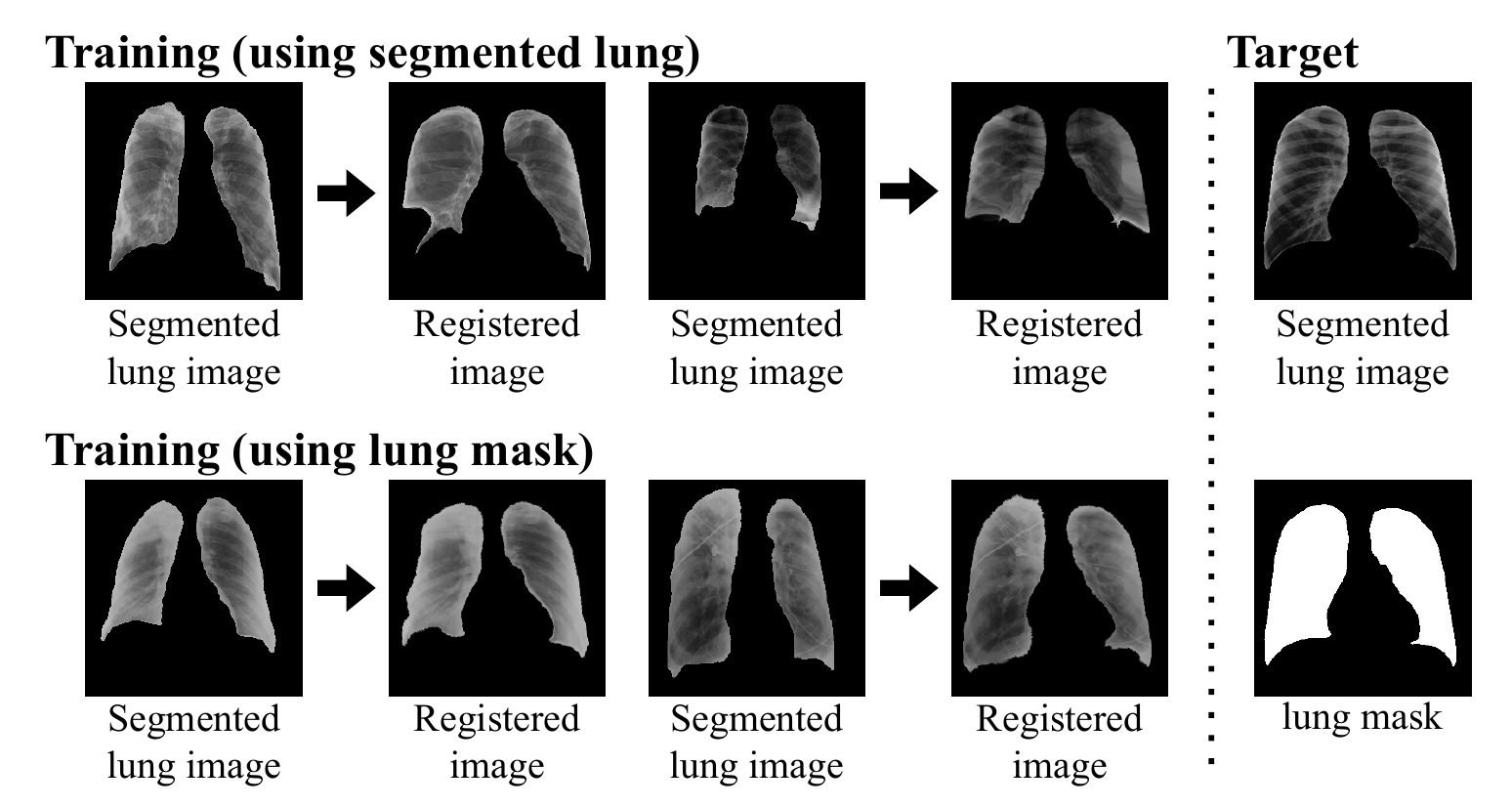}
  \caption{The results of various registration methods on lung images. Row 1: Results of training with segmented lung. Row 2: Results of training with a lung mask. Each training is conducted to register as the target image.}
  \label{reg_base}
\end{figure} 

\textcolor{black}{However, even with the integration of registration networks, the GAN-IT model still has the problem of relying on unpaired data for training. When learning to register lesion data to normal lung areas, lesion areas are more difficult to register than non-lesion areas \cite{xiao2023deep}. The fundamental reason for this limitation is that data containing lesions are learned from data without lesions. Recently, techniques for registration using only label maps (e.g., lung mask) without data have been proposed \cite{hoffmann2021synthmorph}, however as shown in Fig. \ref{reg_base}, current approaches may still lead to incomplete results containing artifacts. Accurate registration is especially important in AL-CXR studies where pixel-level differences are calculated. Therefore, it is important to use reference data (e.g. pseudo-pair datasets) to perform more accurate registration learning.}

To improve GAN-IT for AL-CXR, we proposed a model-agnostic method called image translation under pseudo-paired registration with bilaterally symmetrical data augmentation (IT-PRBA) and an extension of its registration to DL (IT-DPRBA). Both methods aim to improve the AL-CXR performance through a novel registration technique for pre/postprocessing in conventional GAN-IT models.

The proposed IT-PRBA has a unique and reversible registration without requiring DL network training and generates a coordinate transformation map (i.e., the lung image of a moving patient is fit to a fixed lung region) using only a fixed lung mask. As the proposed registration generates a pseudo-paired (i.e., moving/fixed) dataset with only moving images without applying DL, we call it DL-free pseudo-pair registration (DLF-PR). IT-DPRBA extends IT-PRBA through DL-based registration using a paired dataset obtained by DLF-PR, whereas trained registration instead of DLF-PR is used in IT-PRBA. The DL-based method is called DL-based pseudo-paired registration (DL-PR). Unlike existing DL-based registration techniques that simply learn an unpaired dataset \cite{balakrishnan2019voxelmorph,hoffmann2021synthmorph}, the registration in the proposed IT-DPRBA learns a virtual and reasonable paired dataset. This is because the obtained pseudo-paired dataset is less affected by artifacts owing to variations in an unpaired dataset, thereby facilitating finetuning of detailed deformable positions under a virtual and reasonable paired domain in a registration network. The advanced DLF-PR and DL-PR allow GAN-IT to learn relations between abnormal and normal data in a unified domain. 

In IT-PRBA and IT-DPRBA, we introduce bilaterally symmetrical data augmentation (BA), which further improves the performance of GAN-IT for AL-CXR by addressing data imbalance in CXR images containing left/right lung lesions. BA converts an image of the left (right) lung into that of the right (left) lung at the pixel level in the standardized/registered domain. It effectively augments disease data by creating a hypothetical patient with a disease on the other side from a patient with a disease in only one of the left and right lungs. BA is easily implemented owing to the reversible nature of the proposed registration techniques. 
 
We evaluated IT-PRBA and IT-DPRBA by integrating them into GAN-IT models, namely, CycleGAN \cite{zhu2017unpaired} and contrastive unpaired translation (CUT) \cite{park2020contrastive}. We observed that the AL-CXR performance (e.g., for tuberculosis or consolidation shadow cases) is considerably improved when applying our proposal.

\begin{table*}[t]
\caption{\footnotesize {Summary of objectives and characteristics of Related Works and our proposed work.}}
	
	\centering
	{
		\resizebox{\linewidth}{!}{
        \renewcommand{\arraystretch}{1.3}
			\begin{tabular}{cccccc}
				\toprule
             
Work &Backbone& \stackunder{Anomaly}{detection}& {\stackunder{(Q1) Goal is to propose training data translation}{method from unpaired to pseudo-paired ones?}}&{\stackunder{(Q2) If yes for (Q1), does it provide both methods}{based on DL-Free (new baseline) and (its advanced) DL?}}&{\stackunder{(Q3) Goal is to propose a model-agnostic}{method for training the target network?}}\\

				\midrule
     
Schlegl et al. \cite{schlegl2017unsupervised}      &GAN   &Yes& No& No& No\\
Schlegl et al. \cite{schlegl2019f}                 &GAN   &Yes& No& No& No\\
Tang et al. \cite{tang2019abnormal}                &GAN   &Yes& No& No& No\\ 
Zaheer et al. \cite{zaheer2020old}                 &GAN   &Yes& No& No& No\\  
Bhatt et al. \cite{bhatt2021unsupervised}          &GAN   &Yes& No& No& No\\
Keshavamurthy et al. \cite{keshavamurthy2021weakly}&GAN   &Yes& No& No& No\\
Tang et al. \cite{tang2021disentangled}            &GAN   &Yes& No& No& No\\
Zhu et al. \cite{zhu2017unpaired}                  &GAN-IT& No& No& No& No \\
Siddiquee et al. \cite{siddiquee2019learning}      &GAN-IT&Yes& No& No& No \\
Wolleb et al. \cite{wolleb2020descargan}           &GAN-IT&Yes& No& No& No\\
Baur et al. \cite{baur2020steganomaly}             &GAN-IT&Yes& No& No& No \\
Park et al. \cite{park2020contrastive}             &GAN-IT& No& No& No& No\\
Han et al. \cite{han2021dual}                      &GAN-IT& No& No& No& No\\ 
Xu et al. \cite{xu2022maximum}                     &GAN-IT& No& No& No& No\\
Choi et al. \cite{choi2022style}                   &GAN-IT& No& No& No& No\\
Zhan et al. \cite{zhan2022marginal}                &GAN-IT& No& No& No& No\\
Jung et al. \cite{jung2022exploring}               &GAN-IT& No& No& No& No\\
Yang et al. \cite{yang2020mri}                     &GAN-IT& No&Yes& No& No\\
Arar et al. \cite{arar2020unsupervised}            &GAN-IT& No&Yes& No& No\\
Kong et al. \cite{kong2021breaking}                &GAN-IT& No&Yes& No& No\\
Chen et al. \cite{chen2022unsupervised}            &GAN-IT& No&Yes& No& No\\
Ours                                               &GAN-IT&Yes&Yes&Yes&Yes\\

		\bottomrule
   \multicolumn{6}{l}{\footnotesize $\pmb{^*}$Our method is designed to be agnostic for models that utilize a CXR-based image translation backbone.} \\
			\end{tabular}
		}
	}
	\label{tab:relatework}
	
\end{table*}

\section{Related Work}

\subsection{GAN-based Anomaly Localization}

A GAN is a representative DL architecture to synthesize training sets. It can transform unlearned data into samples within learned data. Schlegl et al. \cite{schlegl2017unsupervised} showed that unsupervised learning with GANs using only training data from normal patients allows creating a virtual normal image from the actual image of a patient with anomalies, identifying the abnormal region using the difference between images. Subsequent studies have been conducted to accelerate processing \cite{schlegl2019f}, improve performance \cite{zheng2019one, ngo2019fence, perera2019ocgan}, and apply this method to CXR images \cite{bhatt2021unsupervised,tang2019abnormal}. However, unsupervised learning may generate images with high deviations and unstable anomaly localization \cite{zaheer2020old}.

Weakly supervised GANs based on training with both normal and abnormal labeled data have recently been adopted to further improve performance. In addition to exploiting abnormal data during learning, such GANs outperform those based on unsupervised learning for accurate anomaly localization and stability \cite{keshavamurthy2021weakly, tang2021disentangled}. In principle, these methods are based on image translation \cite{siddiquee2019learning, wolleb2020descargan, baur2020steganomaly}, which synthesizes normal CXR images from CXR images containing disease regions. However, existing GAN- or GAN-IT-based methods for anomaly localization neglect the performance enhancement achievable using registration for learning. In contrast, we explore registration for integration into image translation models as a pre/postprocessing stage aiming to further improve the AL-CXR performance.

\subsection{GAN-based Image Translation}
Various GANs for general image translation have been proposed \cite{zhu2017unpaired,park2020contrastive,han2021dual,xu2022maximum,choi2022style,zhan2022marginal,jung2022exploring}. Zhu et al. \cite{zhu2017unpaired} first demonstrated that image translation can be performed by bidirectionally connecting two GANs in a CycleGAN. Subsequent studies \cite{park2020contrastive, han2021dual, jung2022exploring, zhan2022marginal} addressed the structural imbalance between images before and after translation by enhancing the recognition of spatial location information in image feature maps through contrastive learning. A representative model of this approach is CUT \cite{park2020contrastive}. Another follow-up study on CycleGAN was aimed to enhance the conversion performance by separating style information of an image or applying artificial distortion to an image \cite{xu2022maximum, choi2022style}. Unlike our study, these previous developments neglected registration between unpaired images for learning image translation.

\subsection{Registration-based Image Translation}
Recent GAN-IT models have been developed to improve the IT performance by adopting the deformable registration \cite{yang2020mri,kong2021breaking,chen2022unsupervised,arar2020unsupervised}. Arar et al. \cite{arar2020unsupervised} selectively added a registration network to the input or output of a GAN-IT model, thereby improving the image translation performance by simultaneously learning the translation and registration models. Kong et al. \cite{kong2021breaking} improved the GAN-IT performance by applying data augmentation to introduce random noise for coordinate transformation through registration. Yang et al. \cite{yang2020mri} demonstrated the contribution of registration to GAN-IT for relating different modalities. Chen et al. \cite{chen2022unsupervised} used registration and GAN-IT and removed the GAN discriminator to improve performance. However, unlike our study, the aforementioned existing studies that used registration to IT did not progress registration itself. Specifically, they did not provide an enhanced registration strategy that explicitly addresses the issue of disease artifact fabrication resulting from unpaired data during registration training. In contrast, we use the novel DLF-PR to convert an unpaired dataset into a paired one and proposed DL-PR that can be trained with this virtually paired dataset. This effects as alleviating the artifact generation issue, thereby improving both registration and unpaired IT performance. Overview of related studies is listed in Table \ref{tab:relatework}.

\section{Methods}

\begin{figure}[t]
 \centering
\includegraphics[width=0.7\textwidth]{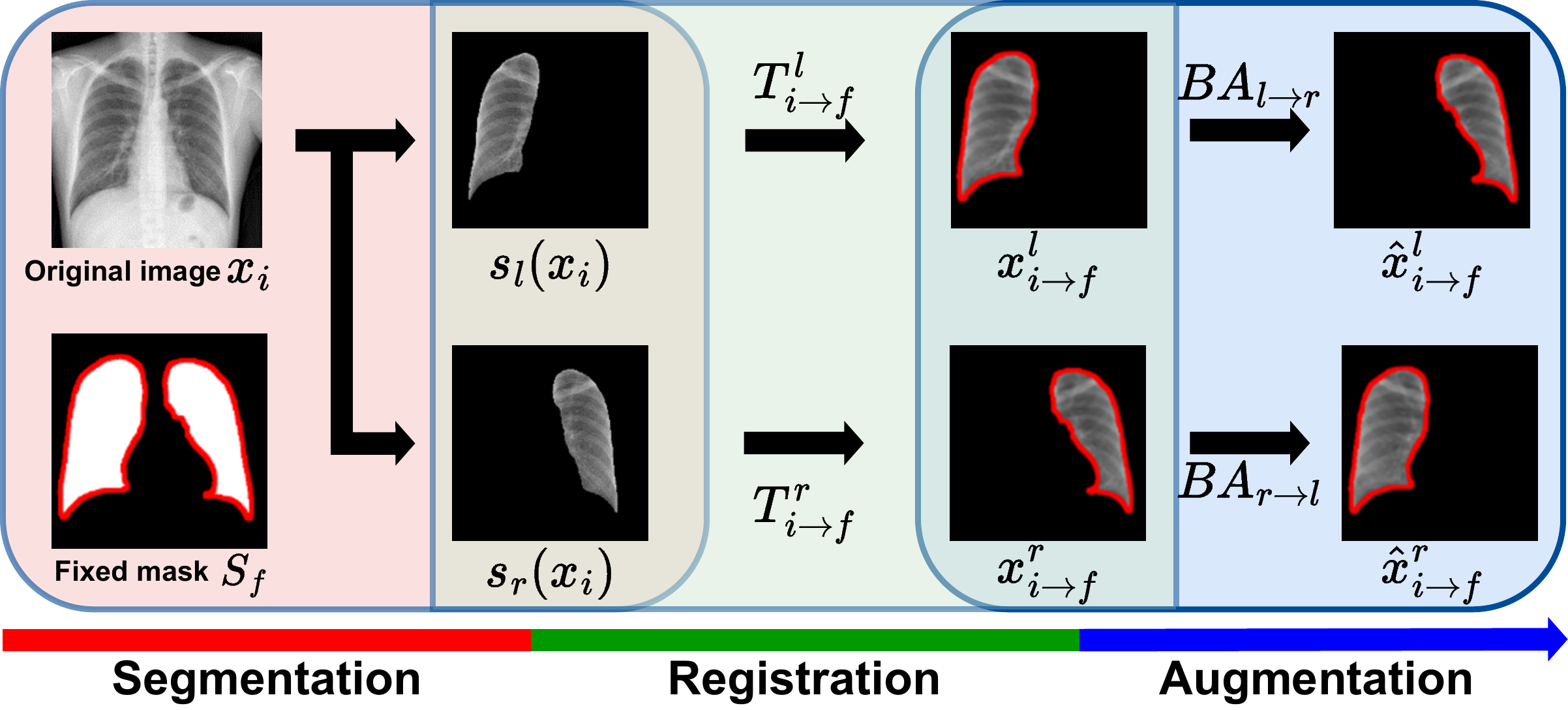}
  \caption{Process of IT-PRBA/IT-DPRBA.}
  \label{overall_bsar_cxr}
\end{figure} 

\subsection{Image translation}

We consider image translation from input domain $X \subset  \mathbb{R}^{h \times w}$ to output domain $Y \subset \mathbb{R}^{h \times w}$. We provide a training dataset for unpaired instances $X = \{ x \in \mathcal{X}\}$ and $Y = \{ y \in \mathcal{Y}\}$. Image translation aims to parameterize mapping $K_{\theta}$ satisfying $K_{\theta} : X \mapsto Y$. This task can be formulated to determine optimal parameter $\theta^*$ of a translation model as follows: 
\begin{align} \label{image_tr_objective} 
\theta^{*} = \argmin_{\theta} \,\, |I(X, \hat{Y}_{\theta}(X)) - I(X, Y)|,
\end{align}
where $I$ denotes the mutual information and $\hat{Y}_{\theta}(X) := \{K_{\theta}(x) $ $\, | \, x \in X \}$ denotes the model translation result for $X$.

\subsection{Baseline GAN-IT for AL-CXR}
\label{bs_int}
Consider $n$ lung CXR images from patients with anomalies and let the indices and dataset be $\{1:n\}:=\{1,2,...,n\}$ and $X = \{x_{i}\}_{i \in \{1:n\}}$, respectively. In addition, consider $m$ lung CXR images from healthy patients with index and datasets denoted as $\{1:m\}$ and $Y = \{y_{i}\}_{i \in \{1:m\}}$, respectively. From the learning data, we train the model using \eqref{image_tr_objective} and perform AL-CXR by identifying anomaly map $v_t$ as the difference between test CXR image $x_t$ (normal or abnormal) and its synthesized result $\hat{y}_t=K_{\theta^*}(x_t)$ as follows:
\begin{align} \label{anomaly_maps} 
  v_t = x_t \odot {\seg}_{\psi^*}(x_t) - \hat{y}_t  \odot {\seg}_{\psi^*}(\hat{y}_t),
\end{align}
where ${\seg}_{\psi^*}$ denotes a pretrained segmentation network that extracts the inner region of the lung as a binary mask. The left- and right-hand terms in \eqref{anomaly_maps} refer to the modified CXR images of $x_t$ and $\hat{y}_t$, respectively, in which the outer lung region is set to zero.

\begin{figure}[t]
 \centering
\includegraphics[width=.95\textwidth]{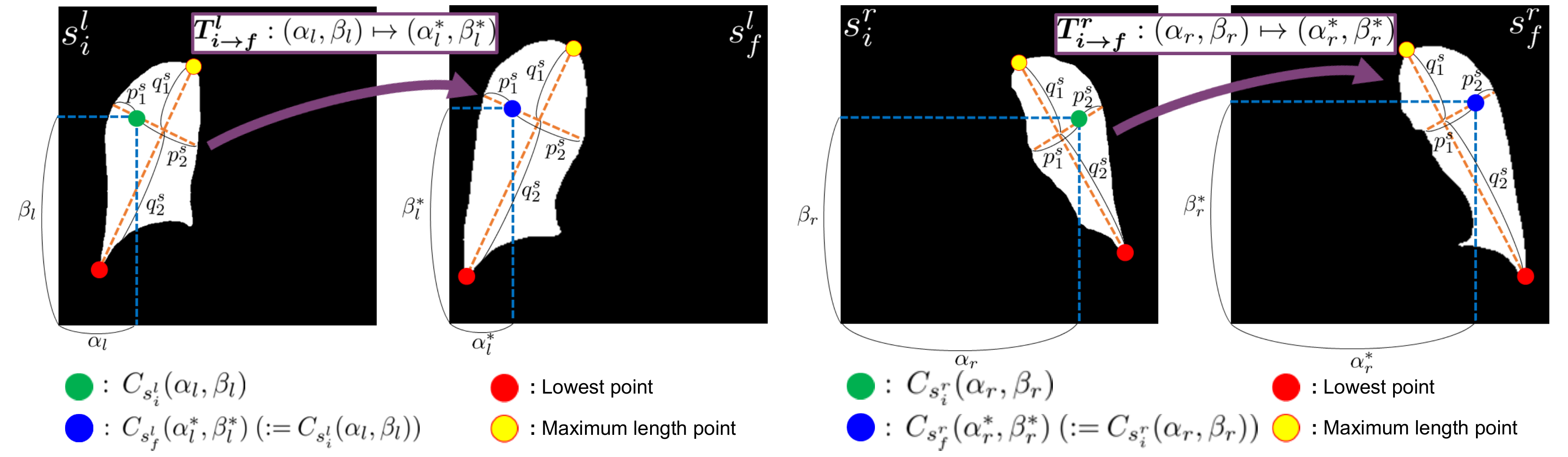}
  \caption{Diagram of proposed DLF-PR process.}
  \label{fig:r_l_C}
\end{figure} 

\begin{figure}[t]
 \centering
   \includegraphics[width=.9\textwidth]{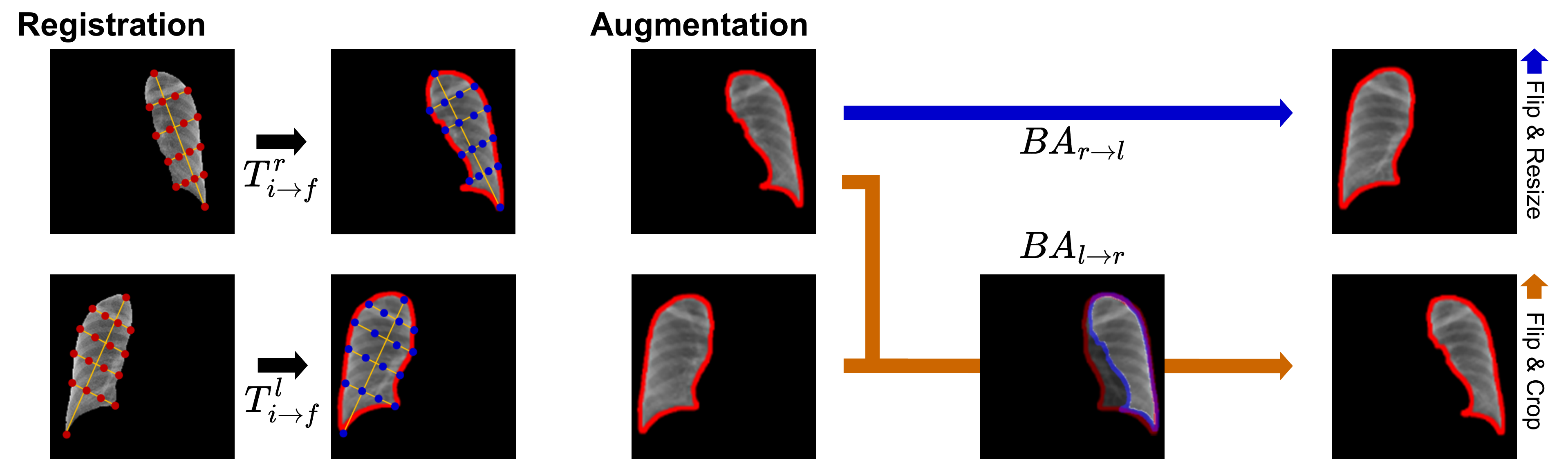}
   \vspace{-0.3cm}
       \caption{Key characteristics of registration and augmentation in IT-PRBA.}
       \label{fig:reg_aug}
\end{figure} 
 \vspace{-0.4cm}

\subsection{Proposed IT-PRBA}
The proposed IT-PRBA (i.e., image translation under pseudo-paired registration with bilaterally symmetrical data augmentation) comprises three stages: segmentation, registration, and BA (i.e., bilaterally symmetrical data augmentation). These stages are described in Algorithm \ref{alg_data_processing} and Fig. \ref{overall_bsar_cxr}.

\subsubsection{Segmentation}
Segmentation of CXR image $x_i$ extracts the left ($s^l_i$) and right ($s^r_i$) lung binary masks and leaves only the region inside the left ($s^l(x_i)$) and right ($s^r(x_i)$) lung images, as shown in Fig. \ref{overall_bsar_cxr}. The binary masks are obtained by splitting the entire mask given as the pretrained segmentation model output, ${\seg}_{\theta^*}(\cdot)$. Then, left ($s^l(x_i)$) and right ($s^r(x_i)$) lung images are obtained by elementwise multiplication $\odot$ of each mask $(s^l,s^r)$ by $x_i$.
\textcolor{black}{A detailed description of the lung segmentation technique we used is provided in Supplementary (Section \ref{segment}).}

{\small\begin{algorithm}[t]
    \begin{small}	
      \caption{IT-PRBA}
      \begin{algorithmic}[1]
      	\SEG
      	\Input{$x_{i}$ for $i \in \{1:n\}$, ${\seg}_{\theta^*}: \mathbb{R}^{h \times w} \mapsto \mathbb{R}^{h \times w}$}
          	\For{$i = 1$ to $n$}
            \State $s_{i} \leftarrow {\seg}_{\psi^*}(x_{i})$
            \State $(s^{l}_{i},s^{r}_{i}) \leftarrow \textup{split}(s_{i})$
            \State $(s^r(x_{i}), s^l(x_{i})) \leftarrow (s^{r}_{i} \odot x_{i}, s^{l}_{i} \odot x_{i})$
            \EndFor
      	\Output{Left and right lung masks $(s^{l}_{i}, s^{r}_{i})$ and their inner images $(s^l(x_{i}), s^r(x_{i}))$  for patient $i$ ($i \in \{1:n\}$)}
      \end{algorithmic}
      \algrule
      \begin{algorithmic} [1]
      	\REG
      	\Input{$x_{i}$, $s^{l}_{i}$, and $s^{r}_{i}$ for $i \in \{1:n\}$}
      	\For{$i = 1$ to $n$}
        \State $(T^l_{i \rightarrow f},T^l_{f \rightarrow i}) \leftarrow {\reg}(s^{l}_{i},s^{l}_{f})$
        \State $(T^r_{i \rightarrow f},T^r_{f \rightarrow i}) \leftarrow {\reg}(s^{r}_{i},s^{r}_{f})$
        \State $(x^l_{i \rightarrow f}, x^r_{i \rightarrow f}) \leftarrow (T^l_{i \rightarrow f}(s^l(x_{i})), T^r_{i \rightarrow f}(s^r(x_{i}))$
      	\EndFor 
      	\Comment{Moving original $(x^r_{i},x^l_{i})$ to reference $(x^r_{f},x^l_{f})$} 
      	\Output{Moved left $x^l_{i \rightarrow f}$ and right $x^r_{i \rightarrow f}$ lung images of patient $i$} 
      \end{algorithmic}
      \algrule
      \begin{algorithmic} [1]
      	\AUG
      	\Input{$x^l_{i \rightarrow f}$, $x^r_{i \rightarrow f}$, $s^{l}_{i}$, and $s^{r}_{i}$ for $i \in \{1:n\}$} 
      	\For{$i = 1$ to $n$}
        \State $(\hat{x}^l_{i \rightarrow f}, \hat{x}^r_{i \rightarrow f}) \leftarrow ({\bsa}_{r \rightarrow l}(x^r_{i \rightarrow f},s^{l}_{f}), {\bsa}_{l \rightarrow r}(x^l_{i \rightarrow f},s^{r}_{f}))$ 
      	\EndFor
      	\Output{Augmented (synthesized) left lung $\hat{x}^l_{i \rightarrow f}$ and right lung $\hat{x}^r_{i \rightarrow f}$ images of patient $i$}
      \end{algorithmic}
      \label{alg_data_processing}
    \end{small} 
\end{algorithm} }

\subsubsection{Registration without DL} We proposed a DL-free registration method and its extension using DL. We first describe DLF-PR; DL-PR is introduced in Section \ref{sec:DL_VM_define}. DLF-PR calculates coordinate shift functions $T^l_{i \rightarrow f}$ and $T^r_{i \rightarrow f}$ mapping from left ($s^l_i$) and right ($s^r_i$) lung masks to those of a fixed lung mask $s_f$, where $i$ and $f$ indicate patient $i$ and fixed, respectively. 

To obtain the corresponding maps, we introduce a coordinate domain to convert general coordinate values $(\alpha,\beta) \in \mathbb{R}^2$ of horizontal/vertical axes at a specific position in a CXR image into relative position coordinates in the lung internal region (inside lung mask $s$) as  $C_{s}(\alpha,\beta)\in [0,1]^4$: 

\begin{align} \label{cs}
C_{s}(\alpha,\beta) := 
\Big(\frac{p^s_1}{p^s_1+p^s_2},\frac{p^s_2}{p^s_1+p^s_2},\frac{q^s_1}{q^s_1+q^s_2},\frac{q^s_2}{q^s_1+q^s_2}\Big).
\end{align}

Examples using $C_{s}(\alpha,\beta)$ are shown in Fig. \ref{fig:r_l_C}. We obtain the longest line starting from the lowest (red) point of each left and right lung, set this line as a new vertical axis and define a perpendicular horizontal axis, and transform original coordinates $(\alpha,\beta) \in \mathbb{R}^2$ into coordinates $C_{s}(\alpha,\beta) \in [0,1]^4$, which are described by the relative position inside the lung expressed between 0 and 1 along the new horizontal and vertical axes.

Considering relative coordinate domain $C_{s}(\alpha,\beta)$, we obtain coordinate shift function $T_{i \rightarrow f}$ such that it maps corresponding relative coordinates $C_{s_i}(\alpha,\beta)$ of the moving patient’s lung mask, $s_i$, to be the same as the corresponding relative coordinates, $C_{s_f}(\alpha,\beta)$, of the fixed patient's lung mask, $s_f$. Particularly, the function can be formulated using \eqref{trans_ifl} and \eqref{trans_ifr} for each lung on the left and right sides, respectively:
\begin{align}\label{lif_moving_transform}
T^l_{i \rightarrow f} &:= \big\{(\alpha,\beta)  \mapsto (\alpha_l^*,\beta_l^*) \,|\, \forall (\alpha,\beta) \in \supp(s^l_i) \big\}, \\\label{trans_ifl}
(\alpha_l^*,\beta_l^*) &:= \underset{(\alpha',\beta') \in \supp(s^l_f)}{\argmin} \left\| C_{s^l_i}(\alpha,\beta) - C_{s^l_f}(\alpha',\beta') \right\|,
\end{align}
\begin{align}\label{rif_moving_transform}
T^r_{i \rightarrow f} &:= \big\{(\alpha,\beta)  \mapsto (\alpha_r^*,\beta_r^*) \,|\, \forall (\alpha,\beta) \in \supp(s^r_i) \big\}, \\\label{trans_ifr}
(\alpha_r^*,\beta_r^*) &:= \underset{(\alpha',\beta') \in \supp(s^r_f)}{\argmin} \left\| C_{s^r_i}(\alpha,\beta) - C_{s^r_f}(\alpha',\beta') \right\|,
\end{align}
where $s^l_i$ and $s^r_i$ indicate the left  and right lung binary masks of patient $i$, respectively and $\supp$ denotes the support set (i.e., nonzero index/location set). Therefore, the value of $(\alpha^*,\beta^*)$ corresponds to the position in the reference lung with respect to the coordinates of $(\alpha,\beta)$ (i.e., the relative position inside the lung becomes the same).

Then, we also define the inverse maps of $T^l_{i \rightarrow f}$ and $T^r_{i \rightarrow f}$ as $T^l_{f \rightarrow i}$ and $T^r_{f \rightarrow i}$, respectively. They map transformed coordinates $(\alpha^*,\beta^*)$ onto original coordinates $(\alpha,\beta)$.  The transformation maps ($T_{i \rightarrow f},T_{f \rightarrow i}$) between the lung images of moving and fixed patients can be obtained only from the mask information of the patients, $s_{i}$ and $s_{f}$. 
Hence, the generation of these maps given by \eqref{lif_moving_transform} and \eqref{rif_moving_transform} can be compactly expressed as function REG: 
\begin{align} \nonumber 
(T^l_{i \rightarrow f},T^l_{f \rightarrow i}) \leftarrow {\reg}(s^{l}_{i},s^{l}_{f}), \, (T^r_{i \rightarrow f},T^r_{f \rightarrow i}) \leftarrow {\reg}(s^{r}_{i},s^{r}_{f}).
\end{align}

The role of these maps is illustrated in Fig. \ref{fig:reg_aug}. We find the longest vertical line (e.g., yellow vertical lines in Fig. \ref{fig:reg_aug}) from the lowest point on the outer edge of each lung image and derive a vertical grid that equally divides the vertical line into various intervals. Then, we obtain each horizontal line (e.g., yellow horizontal lines in Fig. \ref{fig:reg_aug}) that includes each vertical grid perpendicular to the vertical line and derive a horizontal grid that equally divides the horizontal line into various intervals. Then, maps $T^l_{i \rightarrow f}$ and $T^r_{i \rightarrow f}$ are obtained by linearly transforming each point on the grid (e.g., red point for registration in Fig. \ref{fig:reg_aug}) into a point (e.g., blue point for registration in Fig. \ref{fig:reg_aug}) obtained by applying the same process to fixed lung mask $s_f$. To fit the registered images to the original images, we obtain inverse coordinate shift maps $T^l_{f \rightarrow i}$ and $T^r_{f \rightarrow i}$. Additional implementation details for REG are provided in Supplementary (Algorithm \ref{reg_algorithm}). %

Through registration, the relative position inside the lungs of all patients is fixed at specific coordinates. This facilitates learning of image translation using a GAN, even on a small dataset. We demonstrate the effectiveness of the proposed registration compared with existing technologies in Sections \ref{sec:DL_VM_define} and \ref{sec:comp-reg}.

\subsubsection{Training data augmentation by BA} 
The proposed BA (i.e., bilaterally symmetrical data augmentation) doubles the number of images by generating opposite (i.e., right and left) lung images from left and right fixed lung images $x^l_{i \rightarrow f}$ and $x^r_{i \rightarrow f}$ as $\hat{x}^l_{i \rightarrow f}$ and $\hat{x}^r_{i \rightarrow f}$, respectively. Because the right lung image has a horizontally narrower lung region owing to the presence of the heart, after flipping the left image to the right, we partially remove the virtual heart region of the flipped image to fit the right lung (i.e., right lung region excluding the heart), as shown in Fig. \ref{fig:reg_aug} (orange arrow). A detailed description of the proposed BA is provided in Supplementary (Algorithm \ref{bsa_algorithm}), and its validation is presented in Supplementary (Section \ref{eff_aug}).

\subsection{Application of IT-PRBA to GAN-IT for AL-CXR}
\label{sec:applicationtoit}
{\small\begin{algorithm}[t]
\begin{small}	
  \caption{Training}
  \begin{algorithmic}[1]
  	\item[$\textbf{Training for baseline}$]
  	\Input{$X = \{x_{i}\}_{i \in \{1:n\}}$, $Y= \{y_{i}\}_{i \in \{1:m\}}$, $K_{\theta}$} 
  	\State $\theta^{*} \leftarrow \argmin_{\theta} \,\, |I(X, \hat{Y}_{\theta}(X)) - I(X, Y)|$
  	\Output{Trained model $K_{\theta^*}$}
  \end{algorithmic}
  \algrule
  \begin{algorithmic}[1]
  	\item[$\textbf{Training for proposed method using IT-PRBA}$]
  	\Input{$X_r, X_l, Y_r, Y_l$ given by \eqref{aug_res}, $K_{\theta_r}$, $K_{\theta_l}$} 
  	\State $\theta_r^{*} \leftarrow \argmin_{\theta_r} \,\, |I(X_r, \hat{Y}_{\theta_r}(X_r)) - I(X_r, Y_r)|$
  	\State $\theta_l^{*} \leftarrow \argmin_{\theta_l} \,\, |I(X_l, \hat{Y}_{\theta_l}(X_l)) - I(X_l, Y_l)|$  	
  	\Output{Trained models $K_{\theta_r^*}$ and $K_{\theta_l^*}$}	
  	
  \end{algorithmic} 
  \label{algorithm_train}
\end{small}
\end{algorithm}}

{\small\begin{algorithm}[hbt!]
  \begin{small}	
  	\caption{Testing}
  	\begin{algorithmic} [1]
  		\item[$\textbf{Test of baseline}$]
  		\Input{Test CXR image $x_{t}$, $K_{\theta^*}$}  
  		\State $\hat{y}_t \leftarrow K_{\theta^*}(x_t)$
  		\Output{Synthesized normal CXR image $\hat{y}_t$ translated from target CXR image $x_t$}
  		
  	\end{algorithmic}
  	\algrule
  	\begin{algorithmic} [1]
  		\item[$\textbf{Testing of proposed method using IT-PRBA}$]
  		\Input{Test CXR image $x_{t}$, $K_{\theta^*_r}$, $K_{\theta^*_l}$} 
        \State $(x^l_{t \rightarrow f},x^r_{t \rightarrow f}) \leftarrow $ perform up to step 4 of registration by replacing $x_i$ with $x_t$
        \State $\hat{y}^l_t \leftarrow T^l_{f \rightarrow t}(s_l(K_{\theta_l^*}(x^l_{t \rightarrow f})))$
        \State $\hat{y}^r_t \leftarrow T^r_{f \rightarrow t}(s_r(K_{\theta_r^*}(x^r_{t \rightarrow f})))$
        \State $\hat{y}_t \leftarrow s^{l}_{t} \odot \hat{y}^l_t + s^{r}_{t} \odot \hat{y}^r_t + (1- s^{l}_{t}) \odot x_t$  
  		\Output{Synthesized normal CXR image $\hat{y}_t$ translated from target CXR image $x_t$}
  	\end{algorithmic}
  	\label{algorithm_test}
  \end{small}
  
\end{algorithm}}

\subsubsection{Training}
Algorithm \ref{algorithm_train} describes model training using a baseline method and the proposed IT-PRBA. We set the baseline as detailed in Section \ref{bs_int}. In place of the input and label data used in the baseline, we use input data $X_l$ and $X_r$ and label data $Y_l$ and $Y_r$ for the left and right lungs, respectively, in the IT-PRBA as follows. From step 2 in the augmentation stage of Algorithm \ref{alg_data_processing}, we obtain 
\begin{align}\nonumber
X_r &= \{s_r({x}^r_{i \rightarrow f})\}_{i \in \{1:n\}} \cup
\{s_r(\hat{x}^r_{i \rightarrow f})\}_{i \in \{1:n\}}, \\\nonumber
X_l &= \{s_l({x}^l_{i \rightarrow f})\}_{i \in \{1:n\}} \cup \{s_l(\hat{x}^l_{i \rightarrow f})\}_{i \in \{1:n\}}, \\\nonumber 
Y_r &= \{s_r(y_{i})\}_{i \in \{1:m\}}, \\\label{aug_res} 
Y_l &= \{s_l(y_{i})\}_{i \in \{1:m\}}.    
\end{align} 
The number of abnormal samples in $X_r$ or $X_l$ is twice that in $X$ after applying the proposed augmentation. Thus, the amount of data can be considerably increased using our method in a clinical environment in which abnormal (e.g., disease) cases are scarce.

\begin{figure}[t]
 \centering
   \includegraphics[width=.7\textwidth]{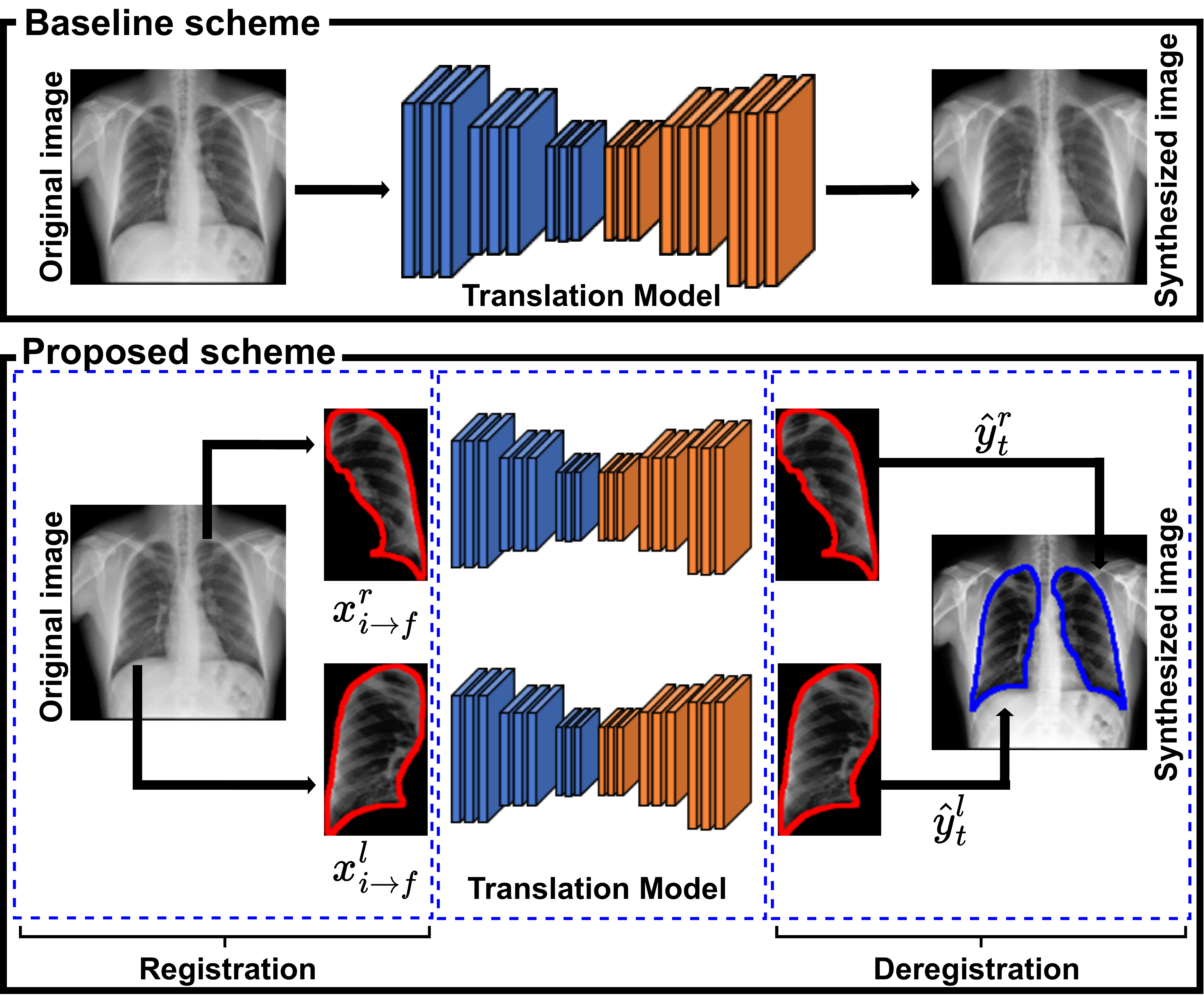} 
        
    	\caption{Testing: proposed IT-PRBA (or IT-DPRBA) compared with baseline GAN-IT.}
    	\label{testing_phase}
       
\end{figure}

\subsubsection{Testing}
\label{sec:testing}

Algorithm \ref{algorithm_test} describes model testing using the baseline and proposed IT-PRBA. Typical baseline image translation methods use original CXR image $x_t$ as the input. However, in the method in which the proposed IT-PRBA is applied, registered image $x_{t \rightarrow f}$ of $x_t$ is used as the input. Subsequently, uniform output $K_{\theta^*}(x_{t \rightarrow f})$ is provided and expressed back in the original image coordinates as $\hat{y}_t = T_{f \rightarrow t}(s_l(K_{\theta^*}(x_{t \rightarrow f})))$. Finally, anomaly map $v_t$ is given by \eqref{anomaly_maps} from $\hat{y}_t$, where $v_t$ represents the difference in the image information between the original image and virtual normal image generated by the network. In other words, the proposed method allows advanced anomaly localization by performing image translation in the registered domain. This process is performed as a separate network (i.e., $K_{\theta_l^*}$ and $K_{\theta_r^*}$) for the left and right lungs (steps 2 and 3), and the results are combined during deregistration (step 4 in Algorithm \ref{algorithm_test}), as illustrated in Fig. \ref{testing_phase}.

\begin{figure*}[t]
\centering
	\includegraphics[width=.8\textwidth]{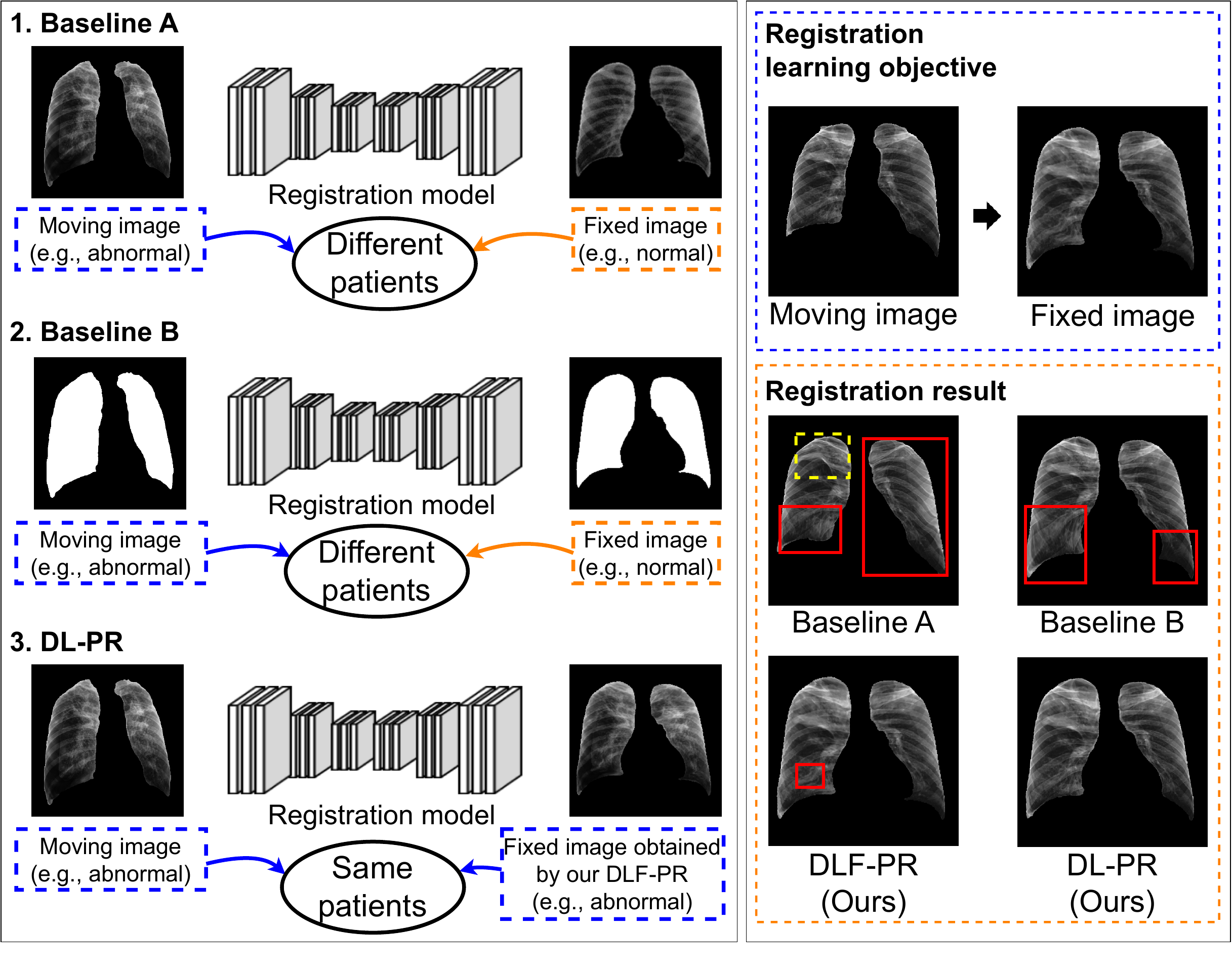}
	\caption{Diagram of proposed DL-PR. Other DL-based baseline techniques (A and B) are unpaired regarding moving and fixed images, impeding proper registration learning (i.e., disease distortion and lung area artifacts as shown in yellow and red boxes respectively). In contrast, DL-PR secures paired (moving/fixed) learning data for the same patient/image source by generating fixed images from moving ones via DLF-PR.}
	\label{method_registration}
 \vspace{-0.2cm}
\end{figure*}

\vspace{-0.2cm}
\subsection{IT-DPRBA: DL-based extension of IT-PRBA}
\label{sec:DL_VM_define}
The proposed DLF-PR performs registration using coordinate transformation function $T_{i \rightarrow f}$ given by \eqref{lif_moving_transform} and \eqref{rif_moving_transform}. The transformation allows the lung region mask of each moving patient to be uniformly mapped onto a reference mask. Hence, a formable instead of deformable coordinate transformation is obtained; however, it cannot achieve perfect lung region coordinate transformation. We address this problem using the proposed DL-PR, which is applicable to any DL-based registration network. Unlike conventional DL-based registration, which uses the learning label as a fixed image, the proposed DL-PR considers a label from a moving image using the DLF-PR output for learning. In other words, existing DL-based registration uses data from different moving and fixed patients, leading to identification of inexistent diseases in images from healthy moving patients and fixed patients with diseases. Conversely, the proposed DL-PR avoids this problem by learning registration using moving and fixed (i.e., result from DLF-PR) images of the same patient. Fig. \ref{method_registration} shows the differences between DL-PR and existing DL-based registration techniques for learning and the different registration performances. For an accurate comparison, the proposed and existing DL-based registration techniques are formulated in detail as follows. 

\paragraph{Proposed DL-based registration technique} For training of the proposed DL-PR, given a registration network $REG_{\phi}$ with training parameter $\phi$, the network takes moving image $x_m$ as its input and learns to output pseudo-label $T_{m \rightarrow f}(x_m)$, which is the moved image generated by DLF-PR. The learning objective is given by  

\begin{small}
\begin{align} \label{proreg_lif} 
\phi^{l*}_{i \rightarrow f} &:= \underset{\phi^l_{i \rightarrow f}}{\argmin} \,\, \mathcal{L}_{\phi^l_{i \rightarrow f}}\Big(REG_{\phi^l_{i \rightarrow f}}(s^l(x_i)), T^l_{i \rightarrow f}(s^l(x_i))\Big), \\\label{proreg_rif} 
\phi^{r*}_{i \rightarrow f} &:= \underset{\phi^r_{i \rightarrow f}}{\argmin} \,\, \mathcal{L}_{\phi^r_{i \rightarrow f}}\Big(REG_{\phi^r_{i \rightarrow f}}(s^r(x_i)), T^r_{i \rightarrow f}(s^r(x_i))\Big),
\end{align}
\end{small}
where $\mathcal{L}_{\phi}(a,b)$ denotes a training loss (e.g., mean squared error) for DL-based registration with network input $a$ (moving image) and label $b$ (fixed image), $s^l(x_i)$ and $s^r(x_i)$ denote the left and right lung images of the moving patient $i$, and $T^l_{i \rightarrow f}(s^l(x_i))$ and $T^r_{i \rightarrow f}(s^r(x_i))$ denote their lung images moved using IT-PRBA. 

From learning described in \eqref{proreg_lif} and \eqref{proreg_rif}, we can obtain smoother/more deformable coordinate transformation functions $\hat{T}^{l}_{i \rightarrow f}$ and $\hat{T}^{r}_{i \rightarrow f}$ (i.e., results by DL-PR) than those provided by DLF-PR (i.e., $T^l_{i \rightarrow f}$ and $T^r_{i \rightarrow f}$); proposed DL-PR calculates coordinate shift functions $\hat{T}^{l}_{i \rightarrow f}$ and $\hat{T}^{r}_{i \rightarrow f}$ mapping from left ($s^l_i$) and right ($s^r_i$) lung masks to those of a fixed lung mask $s_f$ as the output results of following trained registration networks, where $i$ and $f$ indicate patient $i$ and fixed, respectively:
\begin{align} \nonumber 
\hat{T}^{l}_{i \rightarrow f} := REG_{\phi^{l*}_{i \rightarrow f}},  \,\,\,\,\,\,  \hat{T}^{r}_{i \rightarrow f}  := REG_{\phi^{r*}_{i \rightarrow f}},
\end{align}
and DL-PR also provides their inverse maps as $\hat{T}^{l}_{f \rightarrow i}$ and $ \hat{T}^{r}_{f \rightarrow i}$, respectively.

Accordingly by using DL-PR, we proposed an extension of IT-PRBA called IT-DPRBA, which converts all coordinate transformation functions $T^l_{i \rightarrow f}$, $T^r_{i \rightarrow f}$, $T^l_{f \rightarrow i}$, and $T^r_{f \rightarrow i}$ into $\hat{T}^{l}_{i \rightarrow f}$, $\hat{T}^{r}_{i \rightarrow f}$, $\hat{T}^l_{f \rightarrow i}$, and $\hat{T}^r_{f \rightarrow i}$, respectively, in the testing phase of GAN-IT as described in Section \ref{sec:testing} (Algorithm \ref{algorithm_test}). That is, AL-CXR can be performed by applying IT-DPRBA to GAN-IT through the same approach of IT-PRBA by replacing only the coordinate transformation function. This modification from IT-PRBA to IT-DPRBA provides more stable image registration as evaluated in Section \ref{sec:comp-reg}, thereby achieving better anomaly localization performance as shown in Tables \ref{tab:comp_local_inensity_mean} and \ref{tab:comp_local_summask} in Section \ref{al-cxr-comp}.

\paragraph{Existing DL-based registration techniques (Baselines A and B)}  Unlike the DL-PR objective given by \eqref{proreg_lif} and \eqref{proreg_rif}, conventional DL-based registration aims to optimize parameter $\psi$ as $\psi^*$ and obtain coordinate transformation functions $\hat{T}^{l}_{i \rightarrow f}:= REG_{\psi^{l*}_{i \rightarrow f}}$ and $\hat{T}^{r}_{i \rightarrow f}:= REG_{\psi^{r*}_{i \rightarrow f}}$: 

\begin{small}
\begin{align} \label{extreg_lif} 
\psi^{l*}_{i \rightarrow f} &:= \underset{\psi^l_{i \rightarrow f}}{\argmin} \,\, \mathcal{L}_{\psi^l_{i \rightarrow f}}\Big(REG_{\psi^l_{i \rightarrow f}}(s^l(x_i)), s^l(x_f)\Big), \\\label{extreg_rif} 
\psi^{r*}_{i \rightarrow f} &:= \underset{\psi^r_{i \rightarrow f}}{\argmin} \,\, \mathcal{L}_{\psi^r_{i \rightarrow f}}\Big(REG_{\psi^r_{i \rightarrow f}}(s^r(x_i)), s^r(x_f)\Big),
\end{align}
\end{small}
where $s^l(x_i)$ and $s^r(x_i)$ ($s^l(x_f)$ and $s^r(x_f)$) denote the left and right lung images of the moving (fixed) patient, respectively. Training given by \eqref{extreg_lif} and \eqref{extreg_rif} is called DL-based registration baseline A. Baseline A uses image $x_f$ of a patient different from image $x_i$ of a moving patient as a label (i.e., $s^l(x_f)$ and $s^r(x_f)$). This unpaired nature results in unintended disease reduction or distortion artifacts in moved/registered images if the fixed image for training has no disease, but the moving image does. However, our technique described by \eqref{proreg_lif} and \eqref{proreg_rif} uses the label (fixed) image, whose patient information with image content and structure is the same as that of a moving patient image $x_i$. Hence, in DL-PR, the training difference between moving and fixed images exists only for coordinate information, thereby ensuring more reliability for registration learning (i.e., Our DL-PR makes a moving patient being registered to a fixed patient paired with the same moving patient, thereby solving disease reduction or distortion artifacts that baseline A may suffer.).

We also consider another learning objective for DL-based registration in registration baseline B. Unlike baseline A given by \eqref{extreg_lif} and \eqref{extreg_rif}, baseline B uses only binary lung mask information (i.e., $s_i$ and $s_f$) depending on whether it belongs to the lung, without exploiting internal lung information, $s(x_i)$ and $s(x_f)$, of moving and fixed patients, respectively. Baseline B optimizes parameter $\pi$ as $\pi^*$ and obtains coordinate transformation functions $\hat{T}^{l}_{i \rightarrow f}:= REG_{\pi^{l*}_{i \rightarrow f}}$ and $\hat{T}^{r}_{i \rightarrow f}:= REG_{\pi^{r*}_{i \rightarrow f}}$ as follows:

\begin{small}
\begin{align} \label{extreg2_lif} 
\pi^{l*}_{i \rightarrow f} &:= \underset{\pi^l_{i \rightarrow f}}{\argmin} \,\, \mathcal{L}_{\pi^l_{i \rightarrow f}}\Big(REG_{\pi^l_{i \rightarrow f}}(s^l_i), s^l_f\Big), \\\label{extreg2_rif} 
\pi^{r*}_{i \rightarrow f} &:= \underset{\pi^r_{i \rightarrow f}}{\argmin} \,\, \mathcal{L}_{\pi^r_{i \rightarrow f}}\Big(REG_{\pi^r_{i \rightarrow f}}(s^r_i), s^r_f\Big).
\end{align}
\end{small}
In this case of baseline B, unlike baseline A, no artifacts such as anomaly reduction or distortion are generated after registration because the pixel values on the inner region of the lung for both moving and fixed images are the same as 1. However, proper registration fails because the relative position between the coordinates after registration is not uniform, resulting in an unrealistic scenario. This is because during training of baseline B, the intersecting coordinate region, in which the binary masks of the moving and fixed patients are with a value of 1, shows slight motion compared with other regions. 

As shown in Fig. \ref{method_registration} (and Fig. \ref{registration_img}), the proposed DL-PR overcomes limitations of baselines A and B (yellow and red boxes in Figs. \ref{method_registration} and \ref{registration_img} respectively). In fact, DL-PR creates a moving/fixed training pair for the same patient to suppress artifacts influencing baseline A and solves the nonuniform registration of baseline B using information of the lung internal structure rather than a simple binary mask.

\begin{figure*}[hbt!]
\centering
    \subfigure[]{\includegraphics[width=0.85\textwidth]{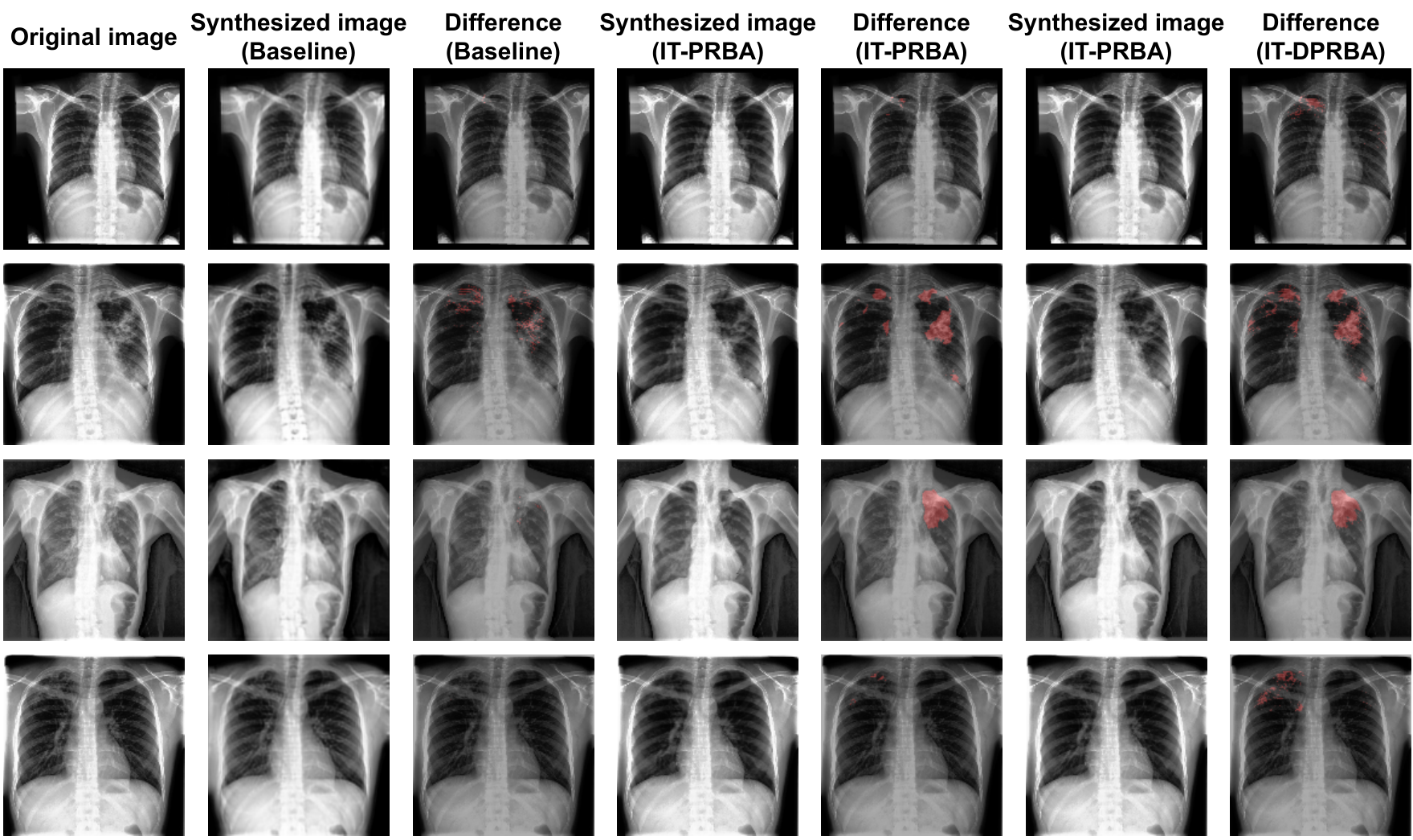}}
    
    \subfigure[]{\includegraphics[width=0.85\textwidth]{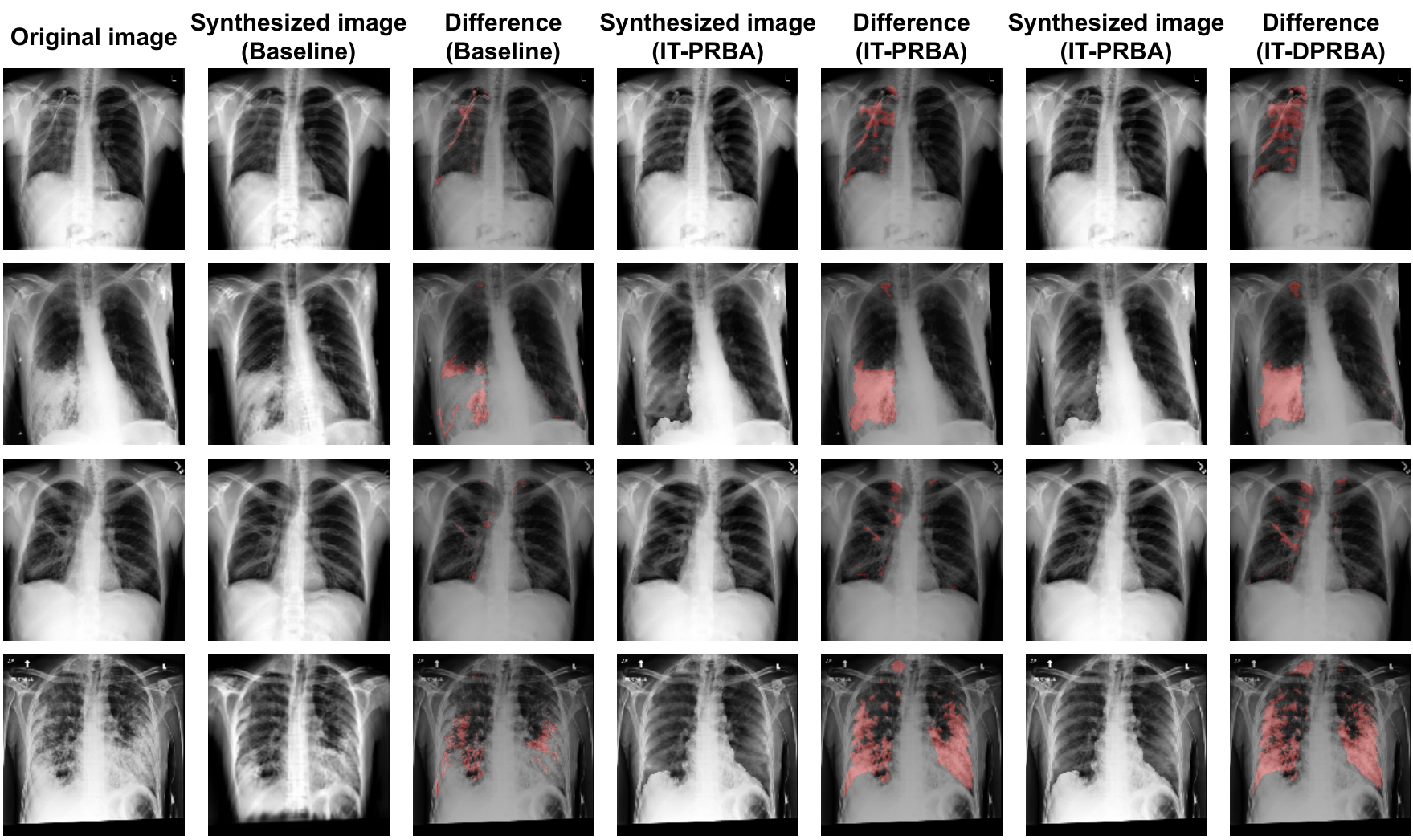}}
    \vspace{-0.3cm}
\caption{Comparison between AL-CXR maps $v_t$ obtained from proposed method and baseline for (a) tuberculosis and (b) consolidation patients. All were commonly used based on CUT.}
\label{fig_comp_CUT}
\vspace{-0.3cm}
\end{figure*}
 
\section{Experiments and Results}

\begin{wraptable}{r}{7.5cm}
	\vskip -15pt 
	\caption{ {Data splitting (in number of samples) for anomaly detection and localization}}
\vskip -10pt
\centering
{
	\resizebox{.95\linewidth}{!}{
		\begin{tabular}{ccccc}
			\toprule
			    Case & Normal& Tuberculosis & Consolidation\\
			\midrule
			 
			Training set & 600 & 600 & 600\\
			 \midrule
			 
			 Test set  & 758 &150 &150\\
			 
			\bottomrule
		\end{tabular}
	}
}
\label{tab:dataset}
 \vspace{-10pt}
\end{wraptable}
\subsection{Data preparation}
We used a publicly available dataset \cite{liu2020rethinking} that provides an anomaly of tuberculosis as a boundary box. From each of the first part of the class sets of abnormal (i.e., tuberculosis) and normal of this data set, we selected $600$ samples for training ($n,m = 600$), and the remaining $150$ and $758$ samples for testing, respectively. In addition to tuberculosis, we collected 750 cases of consolidation patients from publicly available data \cite{liu2020chestx}, of which 600 cases were used for training and the remaining 150 cases for testing.

\subsection{Results for anomaly detection}
\subsubsection{Evaluation Metric: Patient-wise Anomaly Score}
We evaluated AL-CXR using difference $v_t$ in pixel values between the reconstructed and original images given by \eqref{anomaly_maps}. In addition, we calculated the patient-wise anomaly score to obtain a confidence level for whether the target CXR image contains an anomaly. This score was calculated as the $\ell_2$-norm, $\left\| \mathcal{H}(v_t,\tau) \right\|_2$, of anomaly map $v_t$, which was obtained by thresholding, via $\mathcal{H}(v_t,\tau)$, the absolute value of $v_t$, with a value below $\tau \in (20, 30, 40)$ being set to 0 at the pixel level as follows:

\begin{small}
\begin{equation}\label{H_threshold} 
\mathcal{H}_{xy}(v_t,\tau)=
\begin{cases}\nonumber 
  v_{t}(x,y), & \text{if }
       \begin{aligned}\nonumber 
       |v_{t}(x,y)| > \tau  \textup{ for $(x,y) \in \supp(v_t),$}
       \end{aligned}
\\\nonumber 
  0, & \text{for all $(x,y) \notin \supp(v_t).$}
\end{cases}
\end{equation}
\end{small}

\begin{table}[t]

    \centering
    \caption{ AUC of GAN-IT models (CycleGAN and CUT) for patient-wise anomaly score $\left\| \mathcal{H}(v_t,\tau) \right\|_2$ in tuberculosis cases. \textcolor{black}{The receiver operating characteristic curve is presented in Supplementary Fig. S3.}}

	\resizebox{.85\linewidth}{!}{
			\begin{tabular}{c|ccc|ccc}
            \hline
             \multirow{2}{*}{} & \multicolumn{3}{c}{CUT}& \multicolumn{3}{c}{CycleGAN} \\ \cline{2-7} 
             & $\tau = 20$ & $\tau = 30$ &$\tau = 40$ & $\tau = 20$ & $\tau = 30$ & $\tau = 40$ \\  \hline
             Baseline &0.823&0.755&0.691&0.678&0.624&0.586 \\ \hline
				 IT-PRBA (Ours) &0.904&0.910&0.891&0.890 &0.886&0.879 \\ \hline
			 IT-DPRBA (Ours) &$\textbf{0.915}$&$\textbf{0.928*}$&$\textbf{0.919}$&\textbf{0.890}&$\textbf{0.891}$&$\textbf{0.881}$ \\ \hline
			\end{tabular}
		}
	
	\label{tab:result_AUC}
    \vspace{-0.2cm}
    
\end{table}        
\begin{table}[t]
    \centering
    \caption{ AUC of GAN-IT models (CycleGAN and CUT) for patient-wise anomaly score $\left\| \mathcal{H}(v_t,\tau) \right\|_2$  in consolidation cases. \textcolor{black}{The receiver operating characteristic curve is presented in Supplementary Fig. S3.}}
   
    {
	\resizebox{.85\linewidth}{!}{
			\begin{tabular}{c|ccc|ccc}
            \hline
             \multirow{2}{*}{} & \multicolumn{3}{c}{CUT}& \multicolumn{3}{c}{CycleGAN} \\ \cline{2-7} 
             & $\tau = 20$ & $\tau = 30$ &$\tau = 40$ & $\tau = 20$ & $\tau = 30$ & $\tau = 40$ \\  \hline
             Baseline &0.971& 0.964&0.952&0.870& 0.847&0.830 \\ \hline
			 IT-PRBA (Ours)  &{0.975}&0.986&0.984&\textbf{0.975}&0.970&0.960 \\ \hline
			 IT-DPRBA (Ours) &$\textbf{0.982}$&$\textbf{0.991*}$&$\textbf{0.990}$&0.968&$\textbf{0.973}$&$\textbf{0.968}$ \\ \hline
			
			\end{tabular}
		}
	}
	
	\label{tab:result_AUC_col}
\end{table} 
 
To measure the discrimination ability between normal and abnormal cases, we calculated the region under the receiver operating characteristic curve (AUC) of patient-wise anomaly score $\left\| \mathcal{H}(v_t,\tau) \right\|_2$.

\subsubsection{Performance comparison}
\label{sec:al_comp1}
By integrating the proposed IT-PRBA or IT-DPRBA, we trained GAN-IT to evaluate anomaly detection and localization. Anomaly localization was performed on the pretrained GAN-IT model described by \eqref{image_tr_objective} using training data from normal ($Y$) and abnormal ($X$) cases for tuberculosis or consolidation. The results were compared with those of a baseline using an existing GAN-IT without registration and deregistration, as shown in Fig. \ref{testing_phase}. For a fair comparison, we applied the GAN-IT CycleGAN \cite{zhu2017unpaired} or CUT \cite{park2020contrastive} to the proposed and baseline methods. For abnormal data, either tuberculosis or consolidation cases were used. 

Additional details on the training setup of GAN-IT are given as follows: in all the experiments, we adjusted the image size to $256 \times 256$ pixels, set the minibatch size to $24$ and the number of epochs to $150$, and used the Adam optimizer with an initial learning rate of $4 \cdot 10^{-5}$. To optimize performance, the learning rate was linearly decreased until reaching $0$ at the last epoch. We collected CXR unpaired training data corresponding to the input and output of the model to train the GAN-IT model and provided the CXR image of a normal patient by receiving an abnormal CXR image with a specific disease (e.g., tuberculosis \cite{liu2020rethinking} and consolidation \cite{liu2020chestx}) as an input.

The anomaly detection performance for tuberculosis and consolidation is listed in Tables \ref{tab:result_AUC} and \ref{tab:result_AUC_col}, respectively. The AUC of the CUT model is higher than that of CycleGAN in all the cases. For CUT, compared with the existing baseline GAN-IT without IT-PRBA or IT-DPRBA (Fig. \ref{testing_phase}), the proposed IT-PRBA or IT-DPRBA improves the AL-CXR AUCs at various $\tau$ values by more than $4\%$. Among the proposed methods, IT-DPRBA outperforms IT-PRBA in most cases (values in boldface). The values showing the highest performance (marked with *) were obtained from IT-DPRBA. These results validate the proposed IT-PRBA and its DL-based extension, IT-DPRBA. 

Fig. \ref{fig_comp_CUT} shows AL-CXR examples using CUT from localization maps $v_t$ given by \eqref{anomaly_maps} for tuberculosis or consolidation cases. For simplicity, negative values of $v_t$ are treated as 0. The proposed IT-PRBA and IT-DPRBA detect abnormal regions in the lung images that are not detected by the baselines. \textcolor{black}{Examples using CycleGAN for abnormal cases and CycleGAN/CUT for normal cases are shown in Supplementary Figs. S4 and S5.}

\begin{table}[t]
    \centering
    \caption{Comparison of AL-CXR performance using GAN-IT (CUT) between the proposed registration methods (DLF-PR and DL-PR) and other registration methods in tuberculosis and consolidation cases.}
    \resizebox{.85\linewidth}{!}{
			\begin{tabular}{c|ccc|ccc}
            \hline
             \multirow{2}{*}{} & \multicolumn{3}{c}{Tuberculosis}& \multicolumn{3}{c}{Consolidation} \\ \cline{2-7}
             
             & $\tau = 20$ & $\tau = 30$ &$\tau = 40$ & $\tau = 20$ & $\tau = 30$ & $\tau = 40$ \\  \hline
             VoxelMorph \cite{balakrishnan2019voxelmorph} &0.619&0.598&0.580&0.887&0.860&0.830 \\ \hline
        SynthMorph \cite{hoffmann2021synthmorph} & 0.698& 0.679&0.649& 0.828& 0.815& 0.793\\ \hline
	   DLF-PR (Ours) & 0.904& 0.910 & 0.891 & 0.975 & 0.986 & 0.984\\ \hline
        DL-PR (Ours) & \textbf{0.915}& \textbf{0.928*} & \textbf{0.919} & \textbf{0.982} & \textbf{0.991*} & \textbf{0.990}\\ \hline
    
			\end{tabular}
		}
	
	\label{tab:comp_reg_plus}
    \vspace{-0.4cm}
    
\end{table}   
 \subsubsection{Ablation study on the comparison of anomaly detection results using other registration methods.}
\textcolor{black}{In this section, we compare the anomaly detection performance based on our proposed registration methods (i.e., DLF-PR and DL-PR), with other registration methods. For the comparison, we utilized two techniques, VoxelMorph \cite{balakrishnan2019voxelmorph} and SynthMorph \cite{hoffmann2021synthmorph}. All comparative experiments were conducted using the CUT model, which has shown superior performance in GAN-IT, focusing on the cases of tuberculosis and consolidation. The experimental results are listed in Table \ref{tab:comp_reg_plus}. The results show that our proposed methods, DL-PR (AUCs of 0.928 and 0.991) and DLF-PR (AUCs of 0.910 and 0.986), significantly outperform VoxelMorph (AUCs of 0.598 and 0.860) and SynthMorph (AUCs of 0.698 and 0.828) in tuberculosis and consolidation detection, respectively. Through these experiments, we demonstrated the effectiveness of our proposed use of DLF-PR and DL-PR.}

\begin{figure*}[t]
\centering
\subfigure[]{\includegraphics[width=0.32\textwidth]{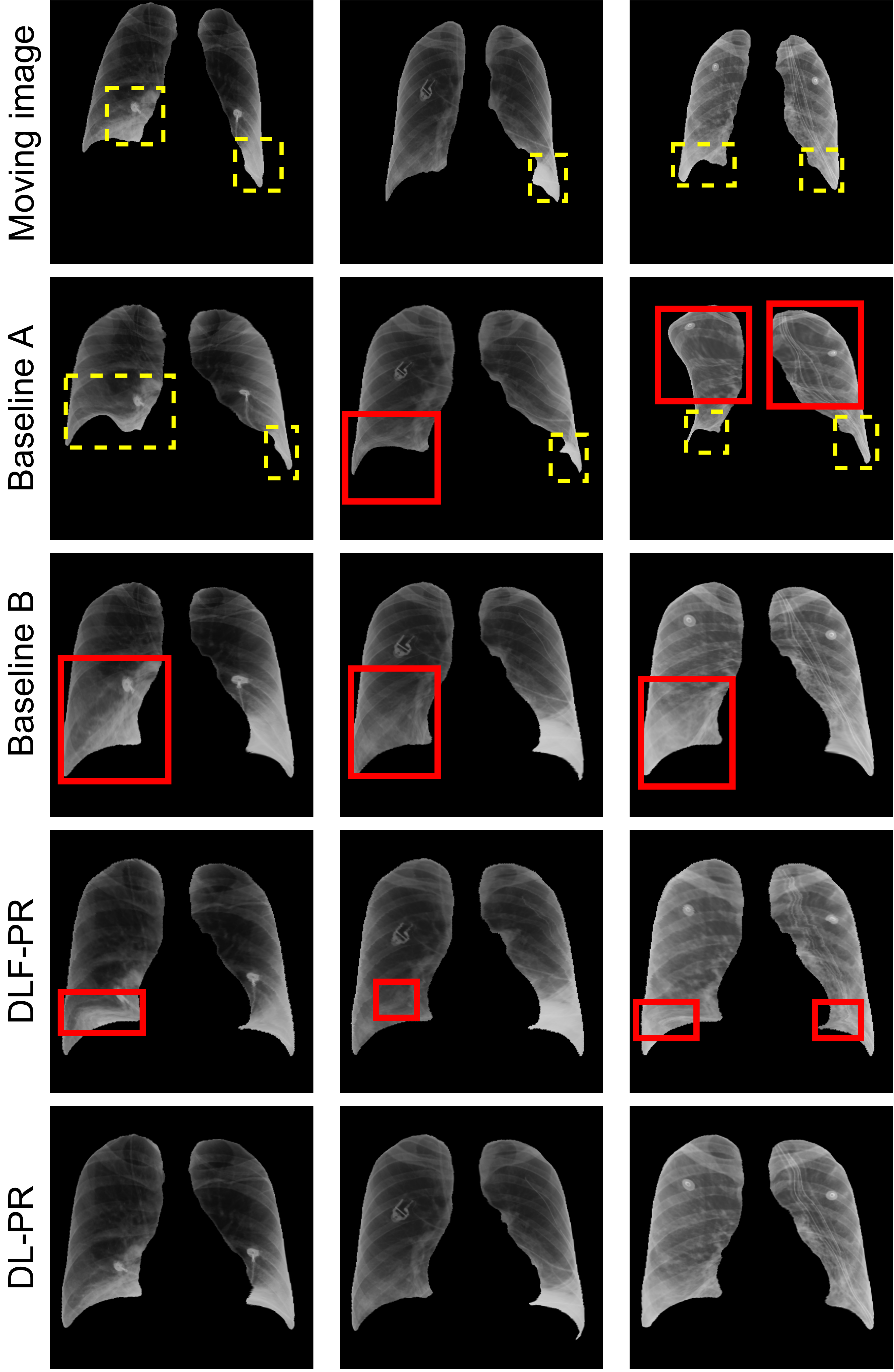}}
   \subfigure[]{\includegraphics[width=0.32\textwidth]{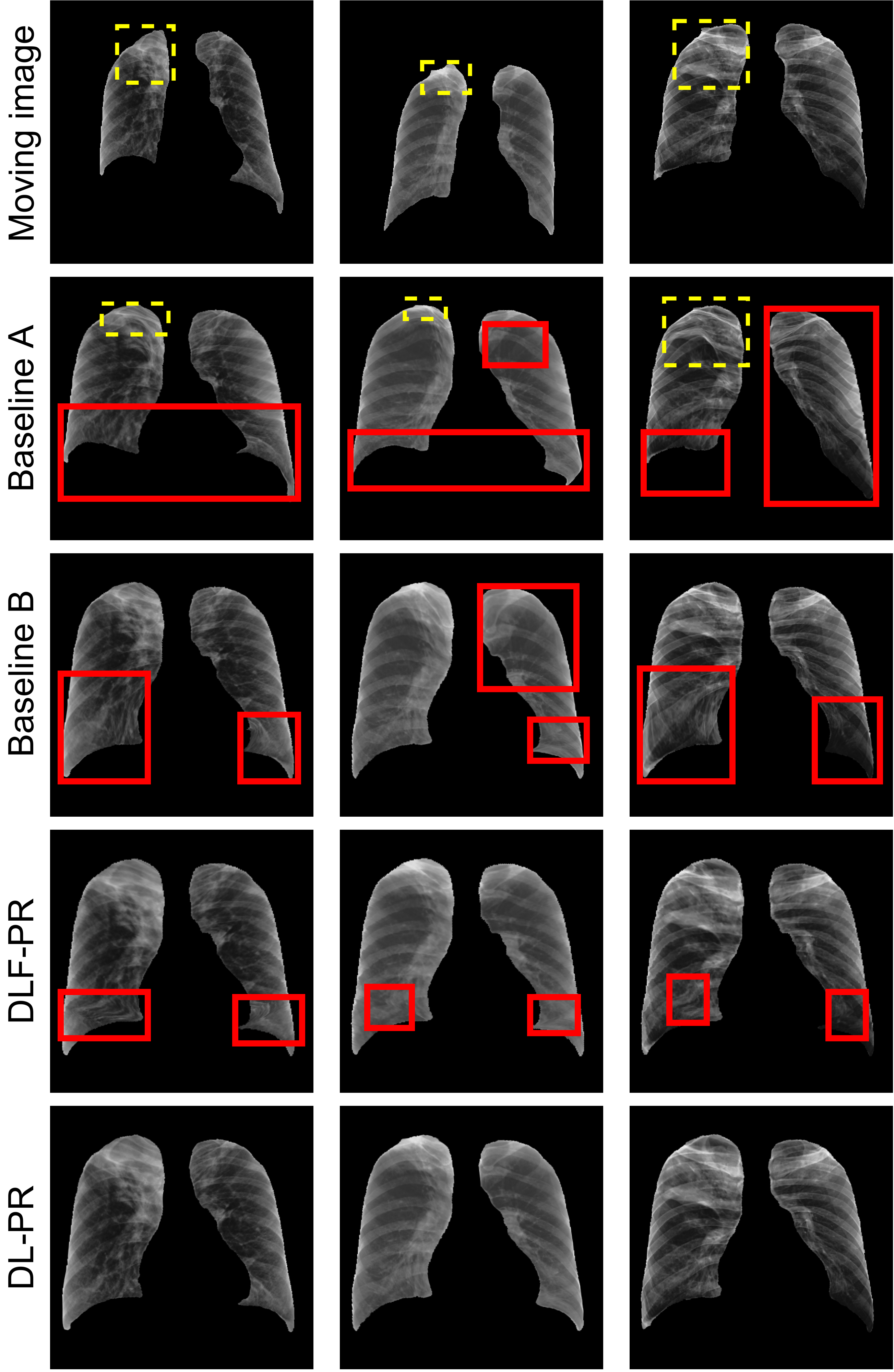}}
  \subfigure[]{\includegraphics[width=0.32\textwidth]{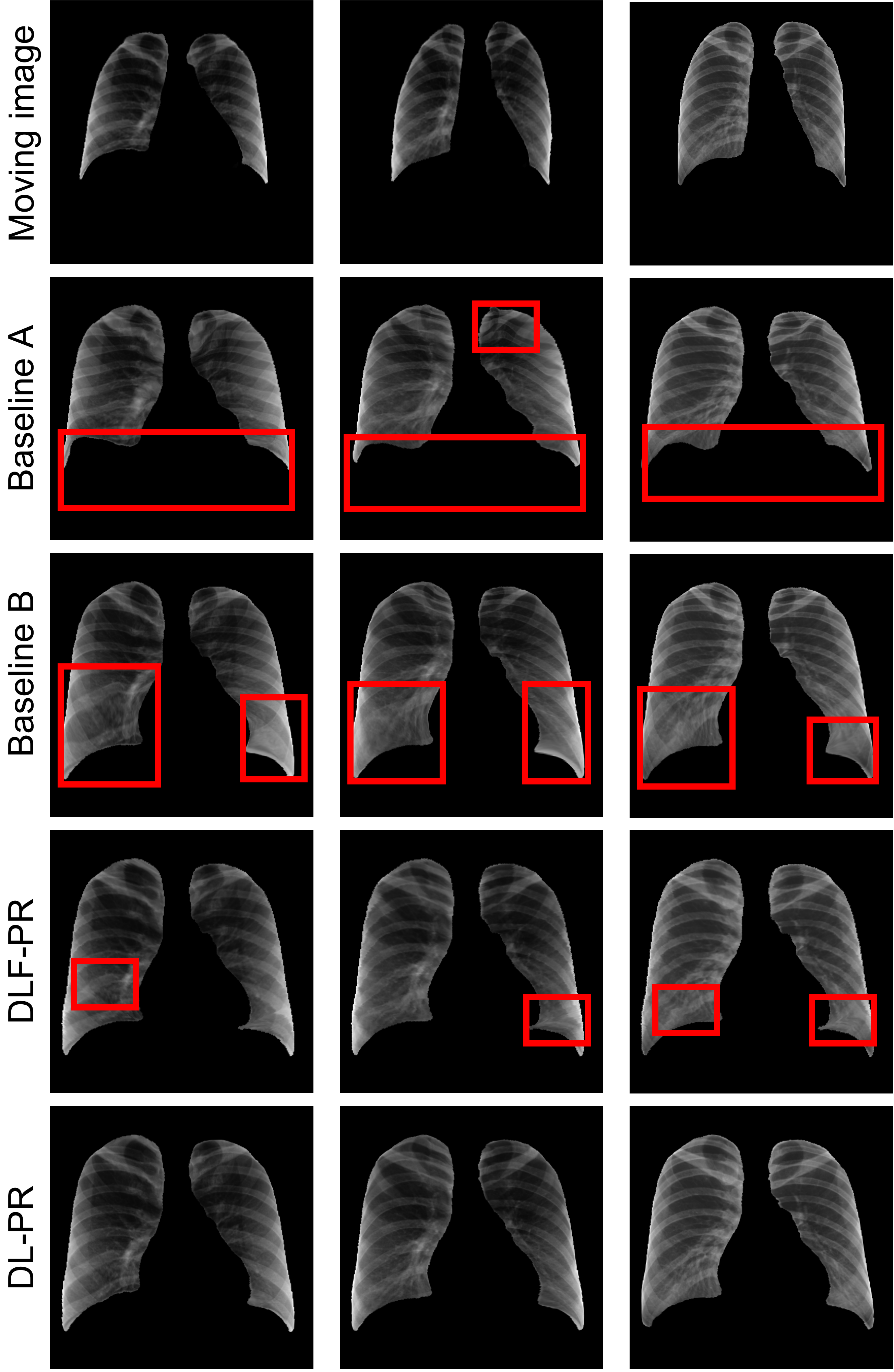}}
  \vspace{-0.3cm}
\caption{Performance of registration techniques for (a) consolidation, (b) tuberculosis, and (c) normal cases. Registered/moved CXR images were individually given by proposed DL-PR/DLF-PR and baselines A/B. Our registration technique provides fewer or no artifacts (Our DL-PR has no disease distortion and reduction (as shown in the yellow box) and artificial shading or insufficient conversion by lung imbalance transformation (as shown in the red box)). Additional images are shown in Supplementary Figs. S7, S8, and S9}
\label{registration_img}
\vspace{-0.2cm}
\end{figure*}

\subsection{Results for anomaly localization}
 
We utilized the public dataset \cite{liu2020rethinking} to demonstrate that the proposed method (i.e., IT-DPRBA and IT-PRBA) provides superior performance for anomaly localization compared to the existing method. Because of the fact that this data set includes tuberculosis location information as a bounding box, model performance can be evaluated by measuring the correlation between the tuberculosis location output by the model and this bounding box.  
\subsubsection{Evaluation metric: Patient-wise anomaly score}
In other words, model performance can be compared by quantifying the proportion of anomaly regions predicted by the proposed model that fall within this bounding box relative to other models. To show this, we represent $z_{t} = v_t \odot g_{t}$ as the intersection area between the anomaly map $v_t$ predicted by the model for the $t$-th patient and the bounding box $g_{t}$, where $g_t \in \{0,1\}^{2}$ denotes the ground truth binary mask for the tuberculosis location bounding box of the $t$-th tuberculosis patient such that the $v_t(x,y)$ value of the pixel location $(x,y)$ in tuberculosis is 1 and otherwise $0$. Then from the intersection area $z_{t}$ of $t$-th patient, we can derive the following two performance evaluation metrics, $s_{intensity}$ and $s_{binary}$ in \eqref{map_score}, for the anomaly localization:  
\begin{align}\nonumber  
s_{intensity}(t) := \sum_{xy}\mathcal{H}_{xy}(z_{t},\tau),\\\label{map_score}
s_{binary}(t) := \sum_{xy}\mathcal{P}_{xy}(z_{t},\tau),
\end{align} 
where  
\begin{equation}\label{P_threshold} 
\mathcal{P}(z_{t},\tau)=
\begin{cases}
  1, & \text{if }
       \begin{aligned}[t]
       z_{t}(x,y) >= \tau  \textup{ for $(x,y) \in \supp(z_{t}),$}
       \end{aligned}
\\
  0, & \text{for all $(x,y) \notin \supp(z_{t}).$}
\end{cases}
\end{equation}
Here, $\mathcal{H}(z_{t},\tau)$ (defined in \eqref{H_threshold}) and $\mathcal{P}_{xy}(z_{t},\tau)$ (defined in \eqref{P_threshold}) refer to the operations delivering the target pixel-wise anomaly score $z_{t}(x,y)$  at location $(x,y)$ as-is or binarizing it respectively if its original score $z_{t}(x,y)$ predicted by the model in the bounding box is greater than a specific threshold. After this operation, we merged the anomaly score value for each pixel inside the bounding box so obtain the final anomaly localization scores $s_{intensity}(t)$ and $s_{binary}(t)$ for each patient (i.e., $t$-th patient). \textcolor{black}{A detailed description of $s_{intensity}(t)$ and $s_{binary}(t)$ is depicted in Supplementary Fig. S6.}

\begin{table}[t]

    \centering
    \caption{ {Comparison of performance between techniques of anomaly  localization in terms of $s_{intensity}$; proposed IT-DPRBA, IT-PRBA, and baseline method ($\tau$ : 25, 30, 35 and the score was divided by $10^3$)}}
    \resizebox{.9\linewidth}{!}{
			\begin{tabular}{c|ccc|ccc}
            \hline
             \multirow{2}{*}{} & \multicolumn{3}{c}{CUT}& \multicolumn{3}{c}{CycleGAN} \\ \cline{2-7}
             
             & $\tau = 25$ & $\tau = 30$ &$\tau = 35$ & $\tau = 25$ & $\tau = 30$ &$\tau = 35$ \\  \hline
             Baseline  & 4.450 & 2.976 & 1.859 & 1.066 & 0.437 & 0.156 \\ \hline
             IT-PRBA (Ours) & 4.984 & 3.641 & 2.610 & 4.370 & 3.334 & 2.517 \\ \hline
		    IT-DPRBA (Ours)  & \textbf{5.822}& \textbf{4.387} & \textbf{3.241} & \textbf{6.361*} &\textbf{4.965}&\textbf{3.821}\\ \hline
			\end{tabular}
		}
	
	\label{tab:comp_local_inensity_mean}

\end{table}
\begin{table}[t]    

    \centering
    \caption{ {Comparison of performance between techniques of anomaly localization in terms of $s_{binary}$; proposed IT-DPRBA, IT-PRBA, and baseline method ($\tau$ : 25, 30, 35 and the score was divided by $10^3$)}}
        \resizebox{.9\linewidth}{!}{
			\begin{tabular}{c|ccc|ccc}
            \hline
             \multirow{2}{*}{} & \multicolumn{3}{c}{CUT}& \multicolumn{3}{c}{CycleGAN} \\ \cline{2-7}
             
             & $\tau = 25$ & $\tau = 30$ &$\tau = 35$ & $\tau = 25$ & $\tau = 30$ &$\tau = 35$ \\  \hline
             Baseline  &156.4&116.7&81.1&33.2&16.4&7.5 \\ \hline
             IT-PRBA (Ours) &194.8&158.0&124.6&180.8&152.4&125.9 \\ \hline
		    IT-DPRBA (Ours)  & \textbf{230.8}& \textbf{192.1} & \textbf{155.6} & \textbf{268.0*} &\textbf{230.5}&\textbf{193.9}\\ \hline
			\end{tabular}
		}
	
		\label{tab:comp_local_summask}
   
\end{table}

\subsubsection{Performance comparison for AL-CXR}
\label{al-cxr-comp}
For these two scores $s_{intensity}$ and $s_{binary}$, we calculated and displayed their mean for a total of $N$ (i.e., $150$) patients in Tables \ref{tab:comp_local_inensity_mean} and \ref{tab:comp_local_summask}, respectively. As a result, it was demonstrated for representative GAN-IT backbone networks (CycleGAN or CUT) and various thresholds $\tau$ that the proposed method (i.e., IT-DPRBA and IT-PRBA) has a consistent performance improvement of AL-CXR compared to the existing baseline method (i.e., an existing GAN-IT without our registration and deregistration). In particular, in the case of the best performance (as marked with $*$ in each of tables), the proposed IT-DPRBA significantly improved the performance of anomaly localization more than $6$ times compared to the baseline. Further, in the comparison between the proposed methods, IT-DPRBA showed a consistent performance improvement compared to IT-PRBA, thereby proving the validity of both proposed methods on AL-CXR. In Fig. \ref{fig_comp_CUT} and Supplementary Fig. S4, the same trend of performance improvement for anomaly localization (i.e., Baseline$<$IT-PRBA$<$IT-DPRBA) can also be observed.

\subsection{Results for deformable registration}
\label{sec:comp-reg}

We compared the proposed registration techniques, namely, DLF-PR and DL-PR, with the existing gold standard DL methods (i.e., Baselines A and B as shown in Section \ref{sec:DL_VM_define}). Accordingly, coordinate transformation functions (from moving to fixed) were individually obtained from DLF-PR using $T^l_{i \rightarrow f}$ in \eqref{lif_moving_transform} and $T^r_{i \rightarrow f}$ in \eqref{rif_moving_transform}, DL-PR using $REG_{\phi^{l*}_{i \rightarrow f}}$ and $REG_{\phi^{r*}_{i \rightarrow f}}$, and DL-based registration baselines A using $REG_{\psi^{l*}_{i \rightarrow f}}$ and $REG_{\psi^{r*}_{i \rightarrow f}}$ and B using $REG_{\pi^{l*}_{i \rightarrow f}}$ and $REG_{\pi^{r*}_{i \rightarrow f}}$, together with each of their inverse maps. 

We used U-Net \cite{isola2017image,ronneberger2015u} as the backbone for DL-based registration because it achieves state-of-the-art performance \cite{balakrishnan2019voxelmorph}. For training, we used mean squared error $\mathcal{L}_{\phi}(a,b)$ as the loss function and the Adam optimizer \cite{kingma2014adam} with a learning rate of 0.0001. The number of epochs was set to 1000, and the batch size was set to 1.

Fig. \ref{registration_img} shows the registration/moved results from the moving images to the fixed images. Unlike other methods, the proposed DL-PR fully registers images without artifacts (no yellow or red box). Relatively few artifacts are observed in DLF-PR compared with the two existing DL-based baselines. As DLF-PR is not based on DL, it does not require training and provides fast registration. DL-PR does not generate artifacts compared with other methods. Therefore, the proposed DLF-PR and DL-PR are effective and achieve high performance.

\section{Conclusion}
Despite the fact that unpaired image translation demonstrates a high potential for quantifying lung disease without pixel annotation, its practical application is limited due to the fundamental issue of performance degradation caused by the inconsistency of data for image translation learning. In other words, the underlying cause of this problem is that the heterogeneity between the reference lung for registration and the moving lung target results in unavoidable artifacts when the image coordinates are converted into one standard region even using the current state-of-the-art DL-based deformable registration techniques. To address this fundamental challenge, we developed a registration technique that employs both linear and (DL-based) non-linear coordinate transformations, enabling consistent and natural movement throughout the lung region. Through this advanced nature, our registration can be applied for the first time as a means for the input and output of an image translation network to result in a single reference domain, whereas previous image translation studies did not apply the registration technique that can unify all images with this single reference domain. In addition, we presented a new data augmentation methodology suitable for this standard domain and applied it in conjunction with the proposed registration to increase the performance of CXR anomaly localization/quantification via image translation in comparison to the prior baselines. Although we only validated our registration-based image translation model on CXR images, its principle can be extended to other applications. Accordingly, we expect our method will serve as a baseline for anomaly localization even without pixel-level annotations of diseases that can be detected using various medical imaging modalities.

\clearpage

\section*{Appendix: Supplementary Materials}

\renewcommand{\thetable}{S\arabic{table}}  
\renewcommand{\thefigure}{S\arabic{figure}}
\renewcommand{\thesection}{S\arabic{section}}

\setcounter{figure}{0} 
\setcounter{table}{0} 
\setcounter{section}{0} 
\setcounter{algorithm}{3}

\section{Implementation details for REG and BA}
\label{appendix:implementation}
In this section, we detail the implementation of the proposed $\reg$ and $\bsa$, which are used in Algorithm 1, as Algorithms \ref{reg_algorithm} and \ref{bsa_algorithm}, respectively. 

\begin{algorithm}[h]
  \begin{small}	
  	\caption{REG}
  	\label{reg_algorithm}
  	\begin{algorithmic} [1]
        \item[$\boldsymbol{T^u_{i \rightarrow f},T^u_{f \rightarrow i} \leftarrow {\reg}(s^{u}_{i},s^{u}_{f}) \,\,\,\, (u \in \{l, r\}):}$ ]
  		\Input{$(s_{i},s_{f})$ for some $i \in \{1:n\}$} 
  		\State ($x^{low}_f,y^{low}_f$) $\leftarrow$ find ($x,y$)-coordinate s.t. $y$ is given as the smallest in boundary set of $s^u_{f}$.
  		\State ($x^{low}_i,y^{low}_i$) $\leftarrow$ find ($x,y$)-coordinate s.t. $y$ is given as the smallest in boundary set of $s^u_{i}$.
  		\State ($x^{high}_f,y^{high}_f$) $\leftarrow$ find ($x,y$)-coordinate s.t. its $l_2$-distance to ($x^{low}_f,y^{low}_f$) is given as the largest in boundary set of $s^u_{f}$.
  		\State ($x^{high}_i,y^{high}_i$) $\leftarrow$ find ($x,y$)-coordinate s.t. its $l_2$-distance to ($x^{low}_i,y^{low}_i$) is given as the largest in boundary set of $s^u_{i}$.
  		\State  $\mathcal{R}_f$ $\leftarrow$ obtain coordinate rotation function $\mathcal{R}_f$ to rotate image $s^u_{f}$ as  $\mathcal{R}_f(s^u_{f})$ s.t. rotated $x^{low}_f$ and rotated $x^{high}_f$ are equal
  		\State  $\mathcal{R}_i$ $\leftarrow$ obtain coordinate rotation function $\mathcal{R}_i$ to rotate image $s^u_{i}$ as  $\mathcal{R}_i(s^u_{i})$ s.t. rotated $x^{low}_i$ and rotated $x^{high}_i$ are equal  
  		\State  $\mathcal{V}_i$ $\leftarrow$ obtain coordinate resizing function $\mathcal{V}_i$ s.t. the longest vertical lengths of positive regions in $\mathcal{V}_i(\mathcal{R}_i(s^u_{i}))$ and $\mathcal{R}_f(s^u_{f})$ are equal and the corresponding columns are on the same $y$-axis coordinates
  		\State  $\mathcal{H}_i$ $\leftarrow$ obtain coordinate resizing function $\mathcal{H}_i$ s.t. for each $y$-axis coordinate, the horizontal lengths of positive regions in $\mathcal{H}_i(\mathcal{V}_i(\mathcal{R}_i(s^u_{i})))$ and $\mathcal{R}_f(s^u_{f})$ are equal 
  		\State  $T^u_{i \rightarrow f}(\cdot)$ $\leftarrow$ obtain the final coordinate transform as $\mathcal{R}^{-1}_f(\mathcal{H}_i(\mathcal{V}_i(\mathcal{R}_i(\cdot))))$, where $\mathcal{R}^{-1}_f(\cdot)$ denotes the inverse map of $\mathcal{R}_f(\cdot)$
   		\State  $T^u_{f \rightarrow i}(\cdot) $ $\leftarrow$ obtain an inverse coordinate transform map of $T^u_{i \rightarrow f}(\cdot)$ s.t. $T^u_{f \rightarrow i}(T^u_{i \rightarrow f}(\cdot))=\mathcal{I}(\cdot)$
  		\Output{$T^u_{i \rightarrow f}, T^u_{f \rightarrow i}$}
  	\end{algorithmic}
  \end{small} 
  \end{algorithm}

\begin{algorithm}[hbt!] 
  \begin{small}	
  	\caption{BA}
  	\label{bsa_algorithm}
  	\begin{algorithmic} [1]
        \item[$\boldsymbol{\hat{x}^l_{i \rightarrow f} \leftarrow {\bsa}_{r \rightarrow l}(x^r_{i \rightarrow f},s^{l}_{f}) :}$  ]
  		\Input{ $(x^r_{i \rightarrow f},s^{l}_{f})$ for some $i \in \{1:n\}$} 
  		 \State  $h^l_{i \rightarrow f}$ $\leftarrow$  flip $\mathcal{R}_f(x^r_{i \rightarrow f})$ horizontally
  		 \State  $h^l_{i \rightarrow f}$  $\leftarrow$  resize vertically $h^l_{i \rightarrow f}$ for its longest vertical line to match that of $\mathcal{R}_f(s^{l}_{f})$
  		 \State  $h^l_{i \rightarrow f}$   $\leftarrow$  resize the horizontal length of $h^l_{i \rightarrow f}$ to match that of $\mathcal{R}_f(s^{l}_{f})$ at  each $y$-axis coordinate
  		 \State  $\mathcal{R}^{-1}_f(\cdot)$ $\leftarrow$  obtain inverse map of $\mathcal{R}_f(\cdot)$ s.t. $\mathcal{R}^{-1}_f(\mathcal{R}_f(s^{l}_{f}))=s^{l}_{f}$
  		\Output{ $\hat{x}^l_{i \rightarrow f} = \mathcal{R}^{-1}_f(h^l_{i \rightarrow f})$} 
   	\end{algorithmic} 
  	\begin{algorithmic} [1]
        \item[$\boldsymbol{\hat{x}^r_{i \rightarrow f} \leftarrow {\bsa}_{l \rightarrow r}(x^l_{i \rightarrow f},s^{r}_{f}) :}$  ]
  		\Input{$(x^l_{i \rightarrow f},s^{r}_{f})$ for some $i \in \{1:n\}$}
  		 \State  $h^r_{i \rightarrow f}$ $\leftarrow$  flip $\mathcal{R}_f(x^l_{i \rightarrow f})$ horizontally
  		 \State  $h^r_{i \rightarrow f}$  $\leftarrow$  resize vertically $h^r_{i \rightarrow f}$ for its longest vertical line to match that of  $\mathcal{R}_f(s^{r}_{f})$
  		 \State  $h^r_{i \rightarrow f}$  $\leftarrow$ remove right-sided/horizontal part of positive regions in $h^r_{i \rightarrow f}$ for remaining part to match that of positive regions in $\mathcal{R}_f(s^r_{f})$ at each $y$-axis coordinate if horizontal length $l_a$ of positive regions in $h^r_{i \rightarrow f}$ is larger than $l_b$ of positive regions in $\mathcal{R}_f(s^r_{f})$ (i.e., $l_a \geq l_b$). Otherwise, resize it for its length to match $l_b$
  		 \State  $\mathcal{R}^{-1}_f(\cdot)$ $\leftarrow$  obtain inverse map of $\mathcal{R}_f(\cdot)$ s.t. $\mathcal{R}^{-1}_f(\mathcal{R}_f(s^{r}_{f}))=s^{r}_{f}$
  		\Output{ $\hat{x}^r_{i \rightarrow f} = \mathcal{R}^{-1}_f(h^r_{i \rightarrow f})$}
  	\end{algorithmic}
  \end{small} 
  \end{algorithm}
  
Registration $\reg$ in Algorithm \ref{reg_algorithm} provides coordinate shift functions $T^l_{i \rightarrow f}$ and $T^r_{i \rightarrow f}$ mapping from left $s^l_i$ and right $s^r_i$ of the lung mask of patient $i$ to $s^l_f$ and $s^r_f$ of a given fixed lung mask $s_f$, respectively, where $i$ and $f$ denote the patient index (for moving images) and fixed image, respectively. Steps 1--4 calculate the longest vertical line (e.g., yellow vertical line in Fig. 3(b)) starting from the lowest point (steps 1 and 2) on the outer edge of each left or right lung mask. Steps 5 and 6 calculate coordinate rotation functions $\mathcal{R}_i$ and $\mathcal{R}_f$, which rotate moving mask $s^u_i$ and fixed mask $s^u_f$, respectively, thus satisfying their longest lines perpendicular to the horizontal axis and calculated in the previous steps. Step 7 calculates vertical coordinate resizing function $\mathcal{V}_i$ for the longest vertical line of the lung in moving $\mathcal{R}_i(s^u_i)$ to match that in fixed $\mathcal{R}_f(s^u_f)$. Step 8 derives horizontal coordinate resizing function $\mathcal{H}_i$ for each horizontal length of the lung image while moving $\mathcal{V}_i(\mathcal{R}_i(s^u_i))$ to match that in fixed $\mathcal{R}_f(s^u_f)$. Finally, step 9 calculates the coordinate transform map as $T^u_{i \rightarrow f}(\cdot)=\mathcal{R}^{-1}_f(\mathcal{H}_i(\mathcal{V}_i(\mathcal{R}_i(\cdot))))$. This includes returning the registered image  $\mathcal{H}_i(\mathcal{V}_i(\mathcal{R}_i(s_i^u)))$ to the original axis of rotation.

The proposed data augmentation algorithms, $\bsa_{r \rightarrow l}$ and $\bsa_{l \rightarrow r}$, in Algorithm \ref{bsa_algorithm} generate additional data by transforming the right ($x^r_{i \rightarrow f}$) and left ($x^l_{i \rightarrow f}$) lung images into left ($\hat{x}^l_{i \rightarrow f}$) and right ($\hat{x}^r_{i \rightarrow f}$) lung images, respectively. Because the right lung image has a horizontally narrower lung region owing to the presence of the heart, after flipping the left image to the right (steps 1 and 2 in ${\bsa}_{r \rightarrow l}$), we partially remove the inner region of the flipped image considering a synthetic heart region to fit the original right lung, as shown in step 3 of ${\bsa}_{r \rightarrow l}$. 

\clearpage

\section{Ablation study for registration and data augmentation in IT-PRBA}
\label{eff_aug}
 
\begin{table}[hbt!]

    \centering
    \caption{ {AL-CXR performance for tuberculosis cases according to whether the proposed DLF-PR and/or BA are applied. B, baseline; R, registration; A, data augmentation; B+R+A, proposed IT-PRBA}}
    \resizebox{.85\linewidth}{!}{
			\begin{tabular}{c|ccc|ccc}
            \hline
             \multirow{2}{*}{} & \multicolumn{3}{c}{CUT}& \multicolumn{3}{c}{CycleGAN} \\ \cline{2-7}
             
             & $\tau = 20$ & $\tau = 30$ &$\tau = 40$ & $\tau = 20$ & $\tau = 30$ & $\tau = 40$ \\  \hline
             B &0.823&0.755&0.691&0.678&0.624&0.586 \\ \hline
             B+R & 0.861& 0.884&0.888& 0.886& 0.878& 0.863\\ \hline
		     B+R+A (Ours)  & \textbf{0.904}& \textbf{0.910} & \textbf{0.891} & \textbf{0.890} &\textbf{0.886}&\textbf{0.879}\\ \hline
			\end{tabular}
		}
	
	\label{tab:comp_aug}
    \vspace{-0.2cm}
\end{table}       
\begin{table}[hbt!]

    \centering
    \caption{ {AL-CXR performance for consolidation cases according to whether the proposed DLF-PR and/or BA are applied. B, baseline; R, registration; A, data augmentation; B+R+A, proposed IT-PRBA}}
        \resizebox{.85\linewidth}{!}{
			\begin{tabular}{c|ccc|ccc}
            \hline
             \multirow{2}{*}{} & \multicolumn{3}{c}{CUT}& \multicolumn{3}{c}{CycleGAN} \\ \cline{2-7}
             & $\tau = 20$ & $\tau = 30$ &$\tau = 40$ & $\tau = 20$ & $\tau = 30$ & $\tau = 40$ \\  \hline
             B  &0.971& 0.964&0.952&0.870& 0.847&0.830 \\ \hline
             B+R          & 0.955& 0.970&0.970&0.975& 0.969& 0.959 \\ \hline
		     B+R+A (Ours) & \textbf{0.975}& \textbf{0.986}&\textbf{0.984}  & \textbf{0.975} & \textbf{0.970} & \textbf{0.960}\\ \hline
			\end{tabular}
		}
	
	\label{tab:comp_aug_con}
\end{table}

\begin{figure}[hbt!]
	\centering

	\subfigure[]{\includegraphics[width=.45\linewidth]{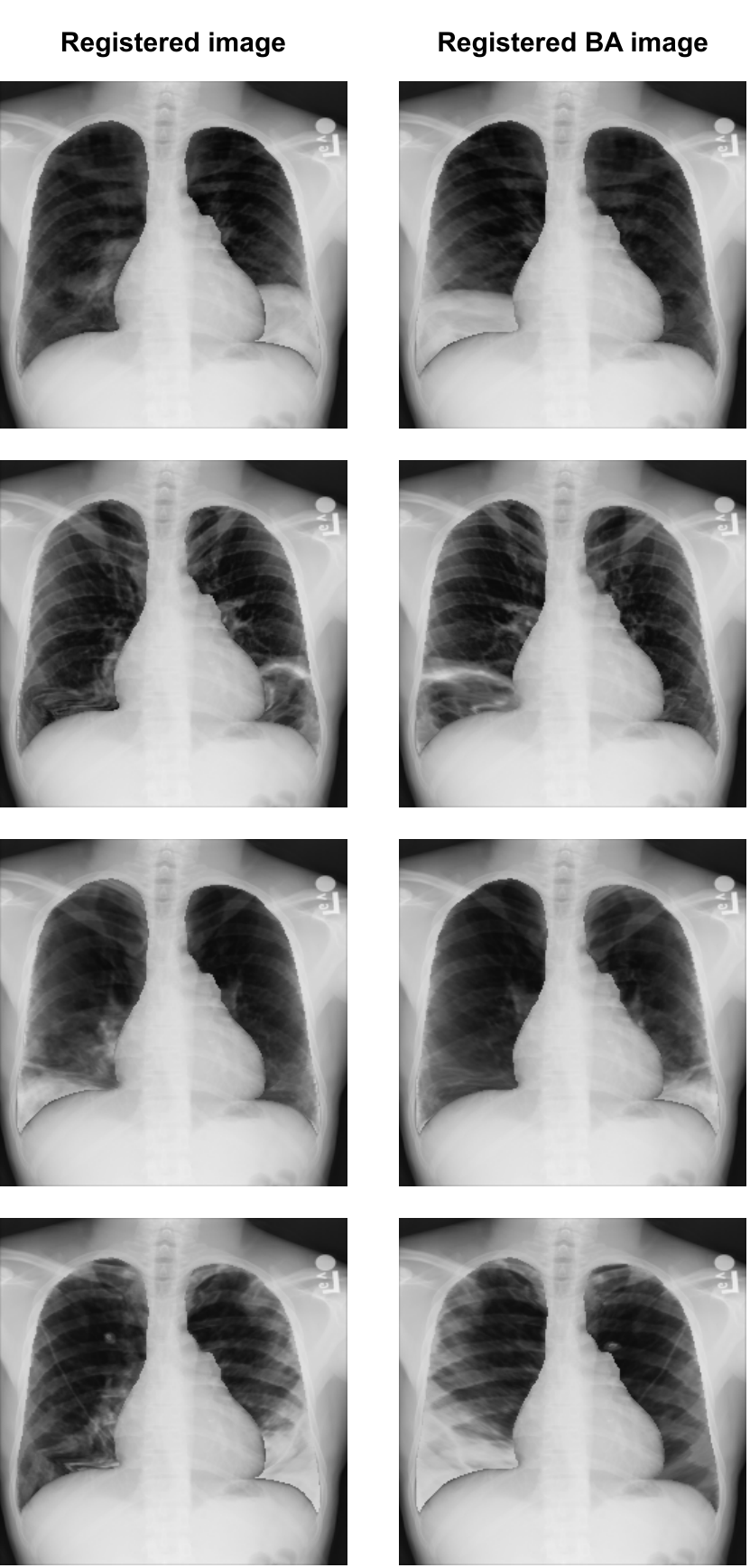}}
	\subfigure[]{\includegraphics[width=.45\linewidth]{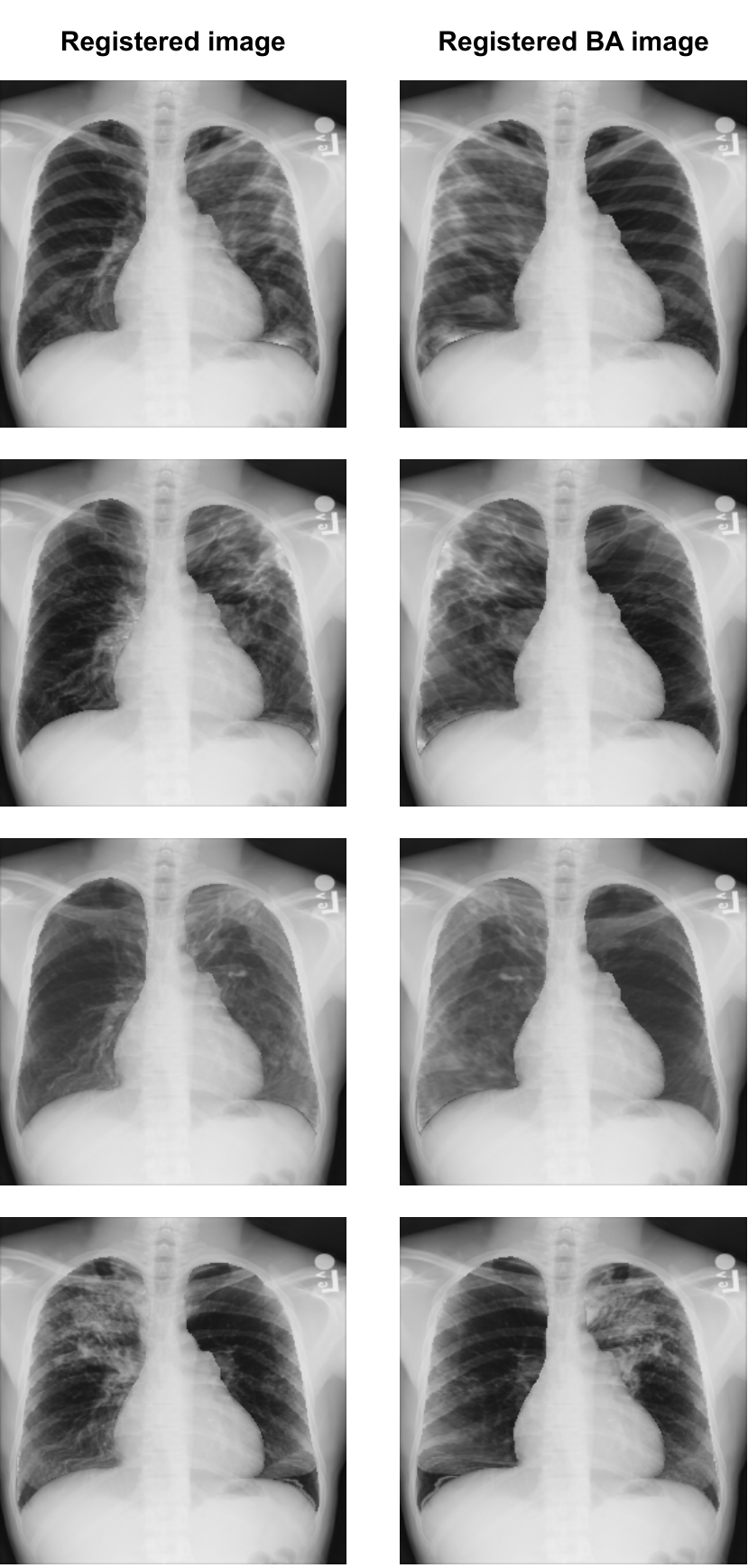}}
	\vskip -8pt
	\caption{Examples of registered and registered BA images for (a) consolidation and (b) tuberculosis cases.} 
	\label{fig:BA_ref_sample}
\end{figure}

\begin{table}[hbt!]
    \centering
    
    \caption{\textcolor{black}{Comparison of AL-CXR performance in tuberculosis and consolidation cases between the proposed BA and other augmentation techniques. B, baseline; R, registration; A, proposed BA; F, flipping; r, rotation; B+R+A, proposed IT-PRBA;}}
    \resizebox{.85\linewidth}{!}{
			\begin{tabular}{c|ccc|ccc}
            \hline
             \multirow{2}{*}{} & \multicolumn{3}{c}{Tuberculosis}& \multicolumn{3}{c}{Consolidation} \\ \cline{2-7}
             
             & $\tau = 20$ & $\tau = 30$ &$\tau = 40$ & $\tau = 20$ & $\tau = 30$ & $\tau = 40$ \\  \hline
             B+R+r (without BA) &\textbf{0.906}&0.885&0.862&\textbf{0.989}&0.985&0.985 \\ \hline
        B+R+F+r (without BA) & 0.826& 0.787&0.775& 0.988& 0.985& \textbf{0.987}\\ \hline
	   B+R+A (with BA) & 0.904& \textbf{0.910}& \textbf{0.891} & 0.975 & \textbf{0.986} & 0.984\\ \hline
      
			\end{tabular}
		}
	
	\label{tab:comp_aug_plus}
    
\end{table}       
The proposed IT-PRBA applies DLF-PR and BA to an existing GAN-IT baseline. We conducted an ablation study to confirm whether DLF-PR and BA contribute to CXR anomaly detection, and the results are listed in Tables \ref{tab:comp_aug} and \ref{tab:comp_aug_con} for tuberculosis and consolidation cases, respectively. These results show that when registration and data augmentation are sequentially added to the baseline, the AL-CXR performance improves, validating IT-PRBA and its components. Additionally, to further demonstrate the contribution of BA to AL-CXR, we conducted comparative experiments with other augmentation techniques (i.e., flipping and rotation). All comparative experiments were conducted using the CUT, which has shown superior performance in GAN-IT, focusing on the cases of tuberculosis and consolidation. The results, listed in Table \ref{tab:comp_aug_plus}, show no significant difference in the consolidation case. However, in the anomaly detection of tuberculosis, we found that the performance of our proposed method, IT-PRBA (AUC 0.910), improved compared to the case of using only image rotation (AUC 0.885) and the case of using both image flipping and rotation (AUC 0.787). Furthermore, through comparisons with various augmentations, we justify the efficacy of using BA for anomaly detection. Examples of BA images for the registered images are shown in Fig. \ref{fig:BA_ref_sample}.

\clearpage
\section{Lung segmentation for data preprocessing}
\label{segment}

\begin{table}[hbt!]
	 
\caption{Segmentation performance in terms of IoU and Dice coefficient of method in \cite{selvan2020lung}}
\vspace{-0.2cm}
\centering
{
	\resizebox{.5\linewidth}{!}{
		\begin{tabular}{ccccc}
			\toprule
			   &\Large{Dataset}  & \Large{IOU} & \Large{Dice}\\
			\midrule
			 \multirow{2}{*}{\Large{Method in \cite{selvan2020lung}}}& \large{JSRT \cite{shiraishi2000development}}  & \large{0.939} & \large{0.969} \\
			 & \large{NLM \cite{jaeger2014two}}  & \large{0.942} & \large{0.970}\\
			\bottomrule
		\end{tabular}
	}
}
\label{tab:segmentation_eval}
\vspace{-0.4cm}
\end{table}

We segmented lung regions from each CXR image using a publicly available pretrained lung segmentation network \cite{selvan2020lung} based on a variational autoencoder, which achieves a higher generalization performance than other networks. We used the JSRT \cite{shiraishi2000development} and NLM \cite{jaeger2014two} datasets to evaluate the lung segmentation performance of the pretrained network in terms of the intersection over union (IoU) and Dice coefficient (DICE) between the actual region of the lung in a 2D image and its prediction, as listed in Table \ref{tab:segmentation_eval}. In both datasets, the values are at least 0.93, confirming a high segmentation performance. Various segmentation results are shown in Supplementary Fig. \ref{fig:segmenation_mask}. 
 IoU (described in (\ref{eval_seg})) and Dice (described in (\ref{eval_seg2})) are obtained through ground-truth (GT) and prediction mask (PM). 
\vspace{-0.1cm}
\begin{align} \label{eval_seg} 
\text{IOU} = \frac{\left| GT \cap PM \right|}{\left| GT \cup PM \right|}\\
\label{eval_seg2} 
\text{DICE} = \frac{2 \left| GT \cap PM \right|}{\left| GT \right| + \left| PM \right|}
\end{align}

\begin{figure}[hbt!]
	\centering

	\subfigure[]{\includegraphics[width=0.18\linewidth]{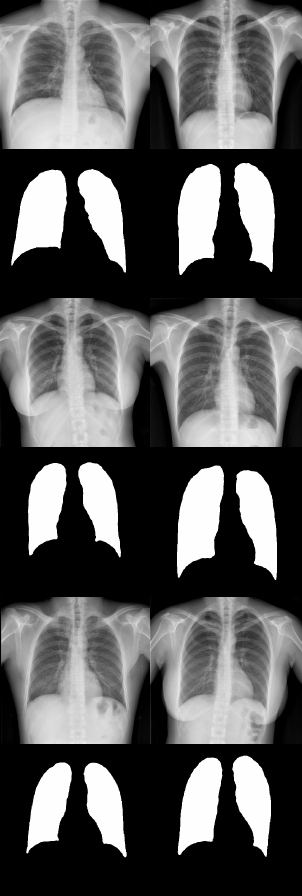}}
	\subfigure[]{\includegraphics[width=0.18\linewidth]{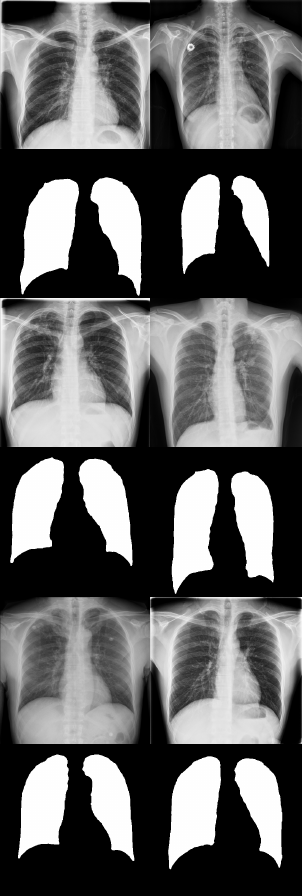}}
	\subfigure[]{\includegraphics[width=0.18\linewidth]{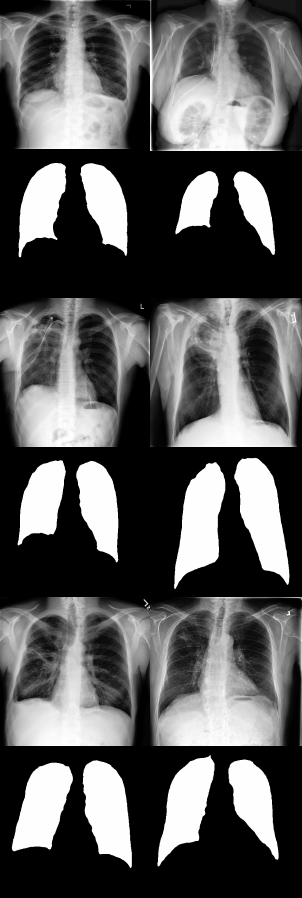}}
	\vskip -8pt
	\caption{ {Examples of lung masks predicted by the segmentation network for (a) normal, (b) tuberculosis, and (c) consolidation cases.}} 
	\label{fig:segmenation_mask}
\end{figure}

\begin{figure*}[h]
\centering
    \subfigure[]{\includegraphics[width=0.9\textwidth]{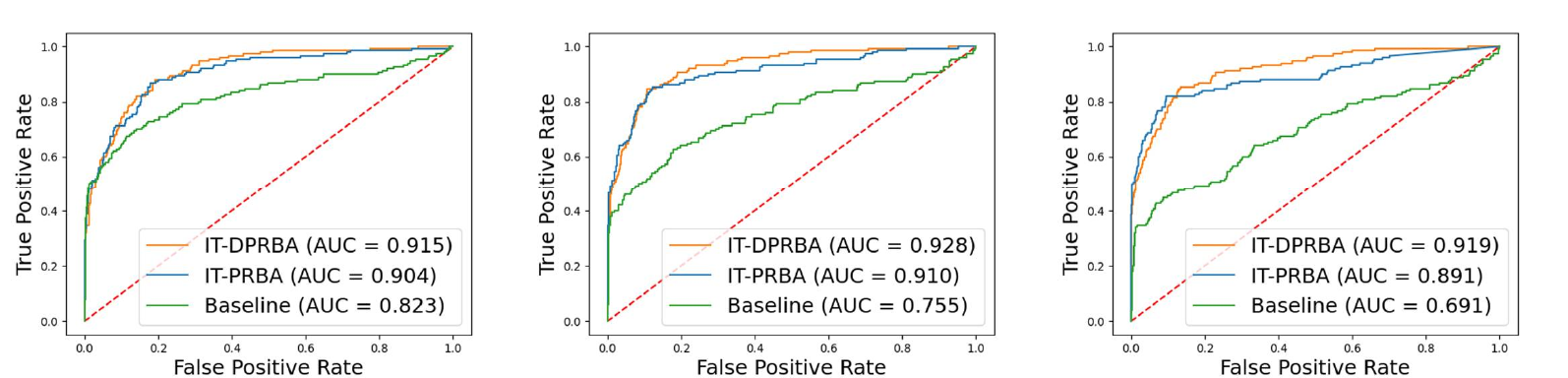}}
    \subfigure[]{\includegraphics[width=0.9\textwidth]{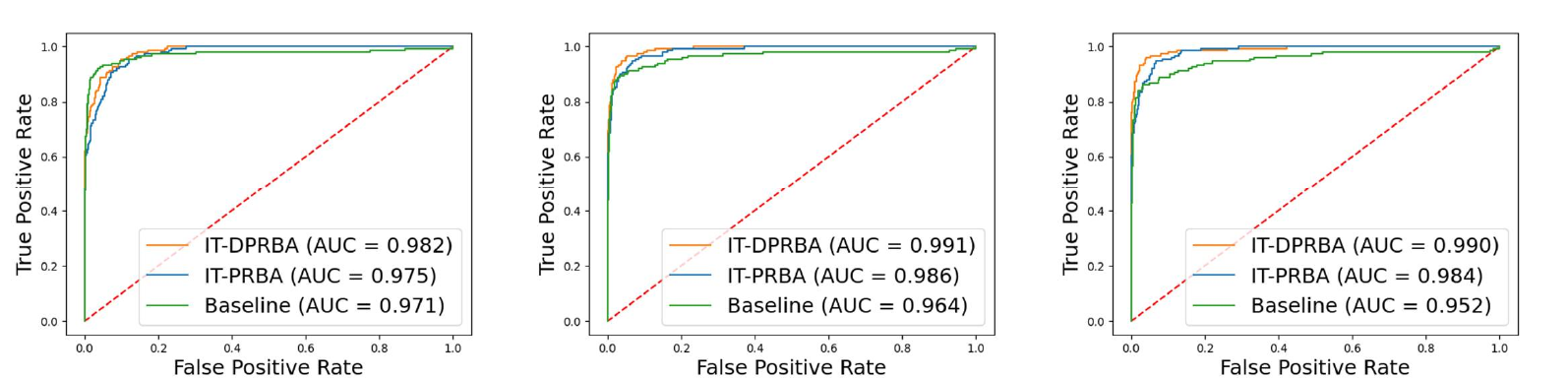}}
    \subfigure[]{\includegraphics[width=0.9\textwidth]{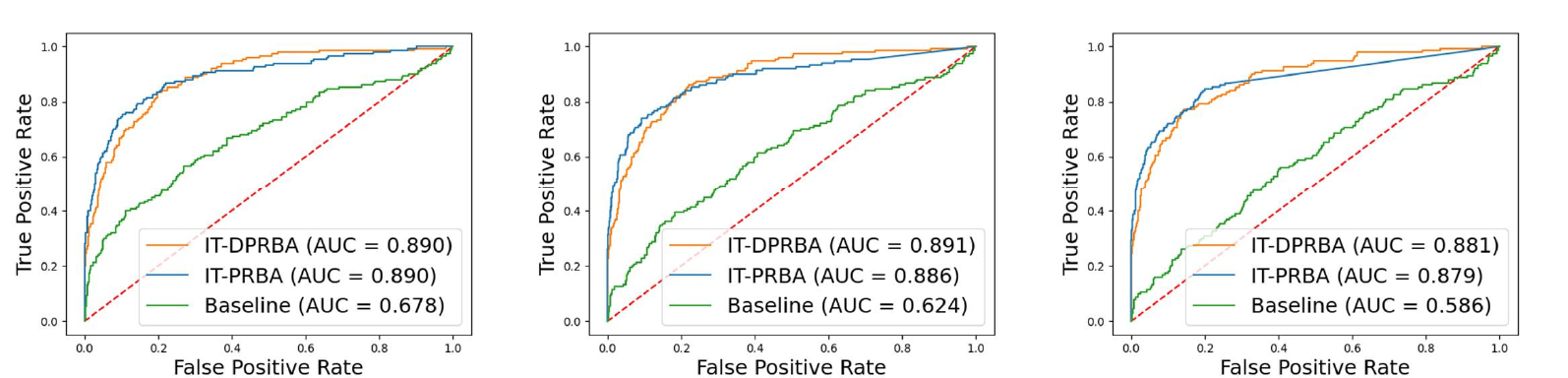}}
    \subfigure[]{\includegraphics[width=0.9\textwidth]{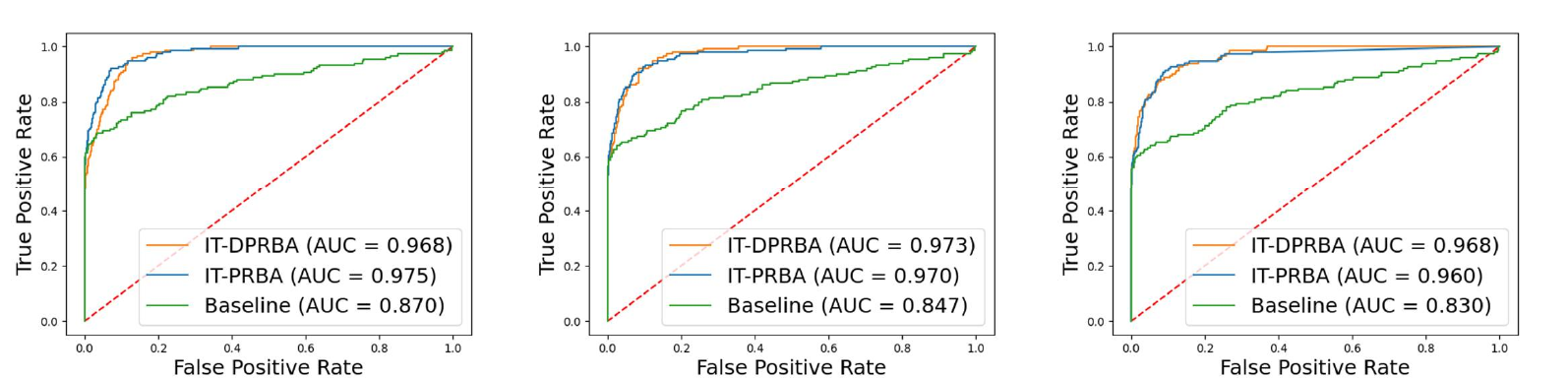}}
\caption{Performance comparison of AL-CXR in terms of AUC with varying threshold $\tau=(20,30,40)$ (from left to right). Proposed method compared with CUT baseline for (a) tuberculosis and (b) consolidation cases. Proposed method compared with CycleGAN baseline for (c) tuberculosis and (d) consolidation cases.}
\label{fig_comp_cut}
\end{figure*}

\begin{figure*}[h]
\centering
    \subfigure[]{\includegraphics[width=0.9\textwidth]{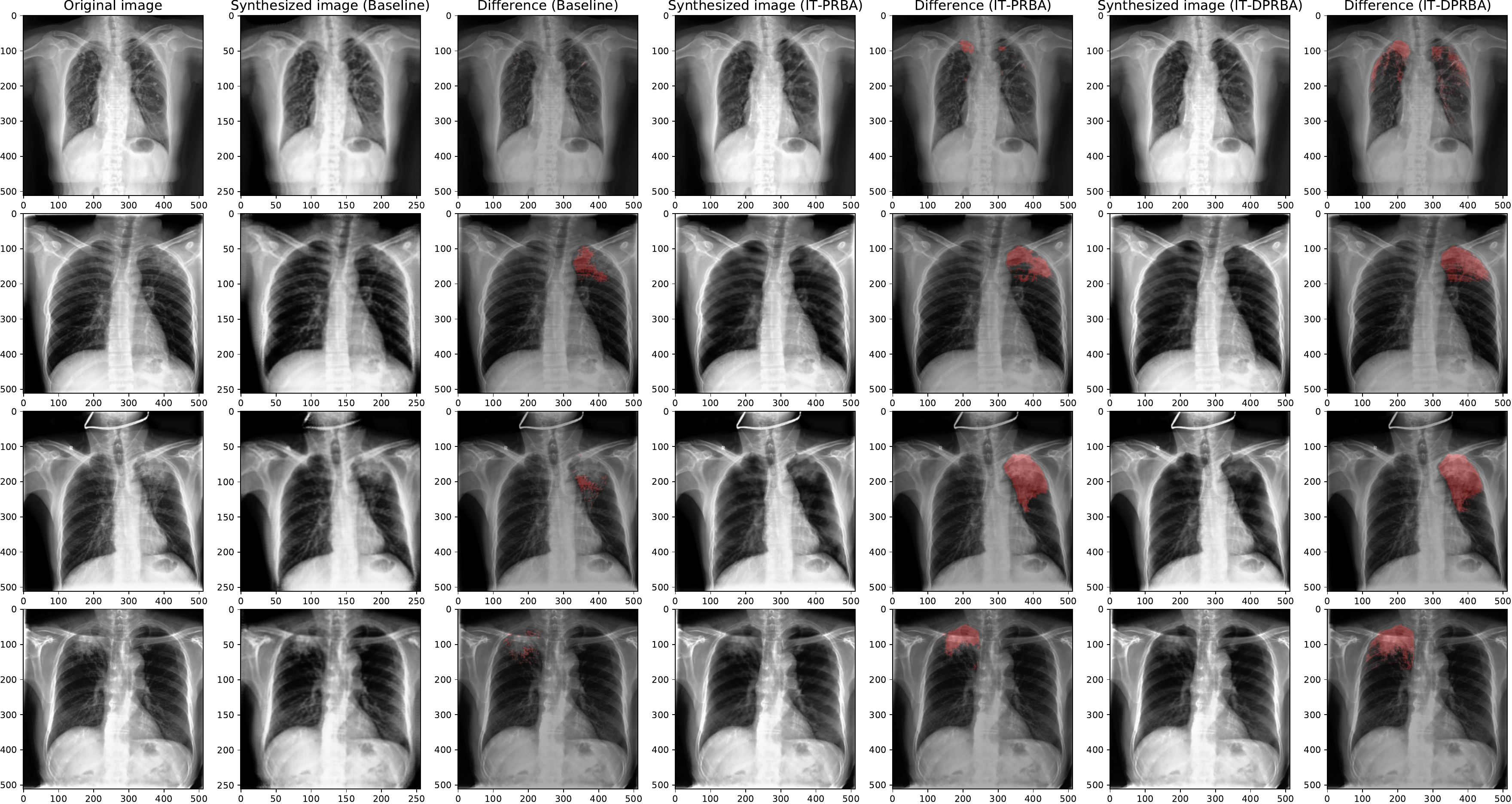}}
    \subfigure[]{\includegraphics[width=0.9\textwidth]{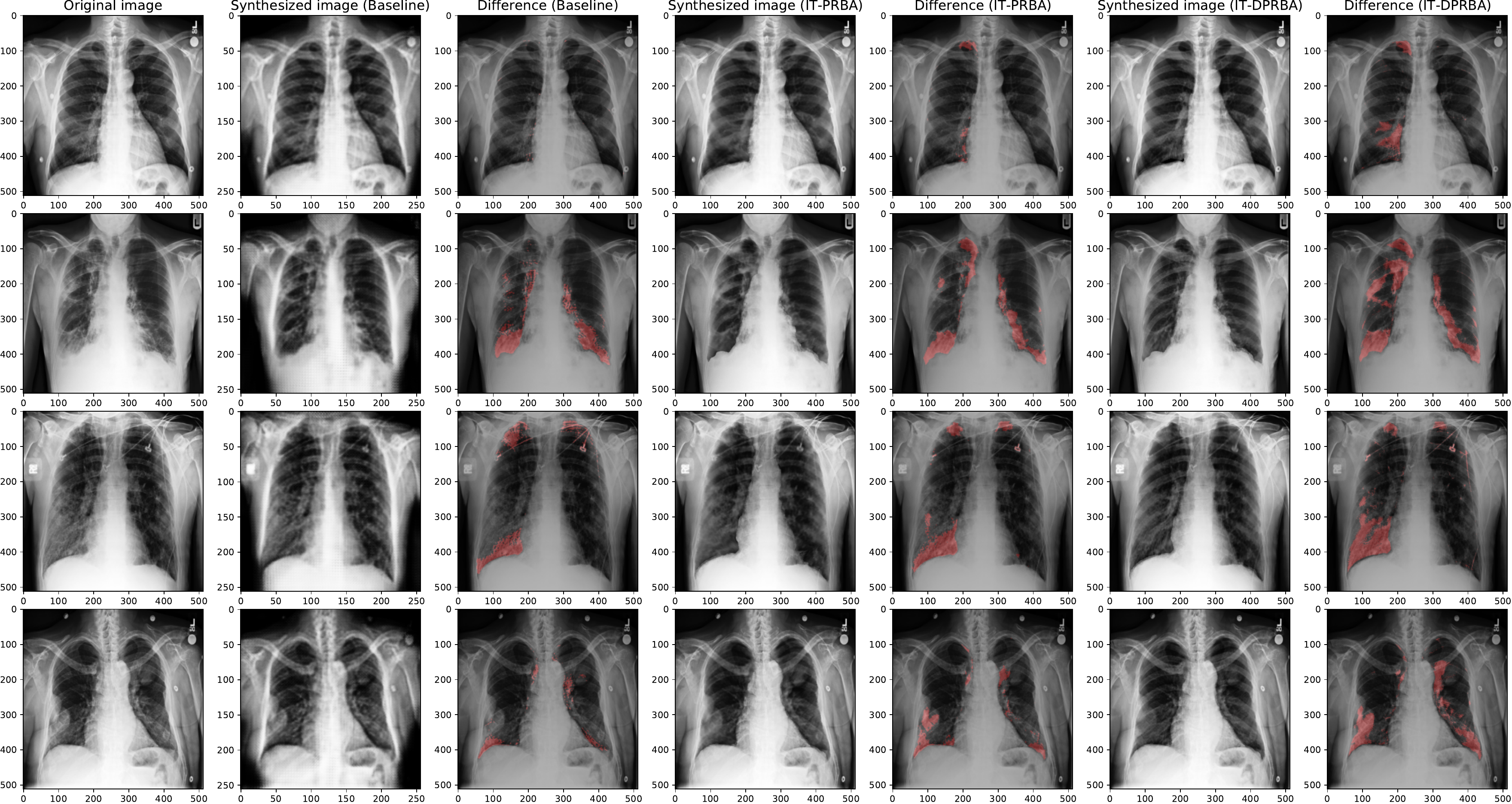}}
\caption{Comparison between anomaly localization maps $v_t$ obtained from proposed method and baseline for (a) tuberculosis and (b) consolidation cases. All were commonly used based on CycleGAN.}
\label{fig_comp_cycle}
\end{figure*}

\begin{figure*}[h]
\centering
    \subfigure[]{\includegraphics[width=0.9\textwidth]{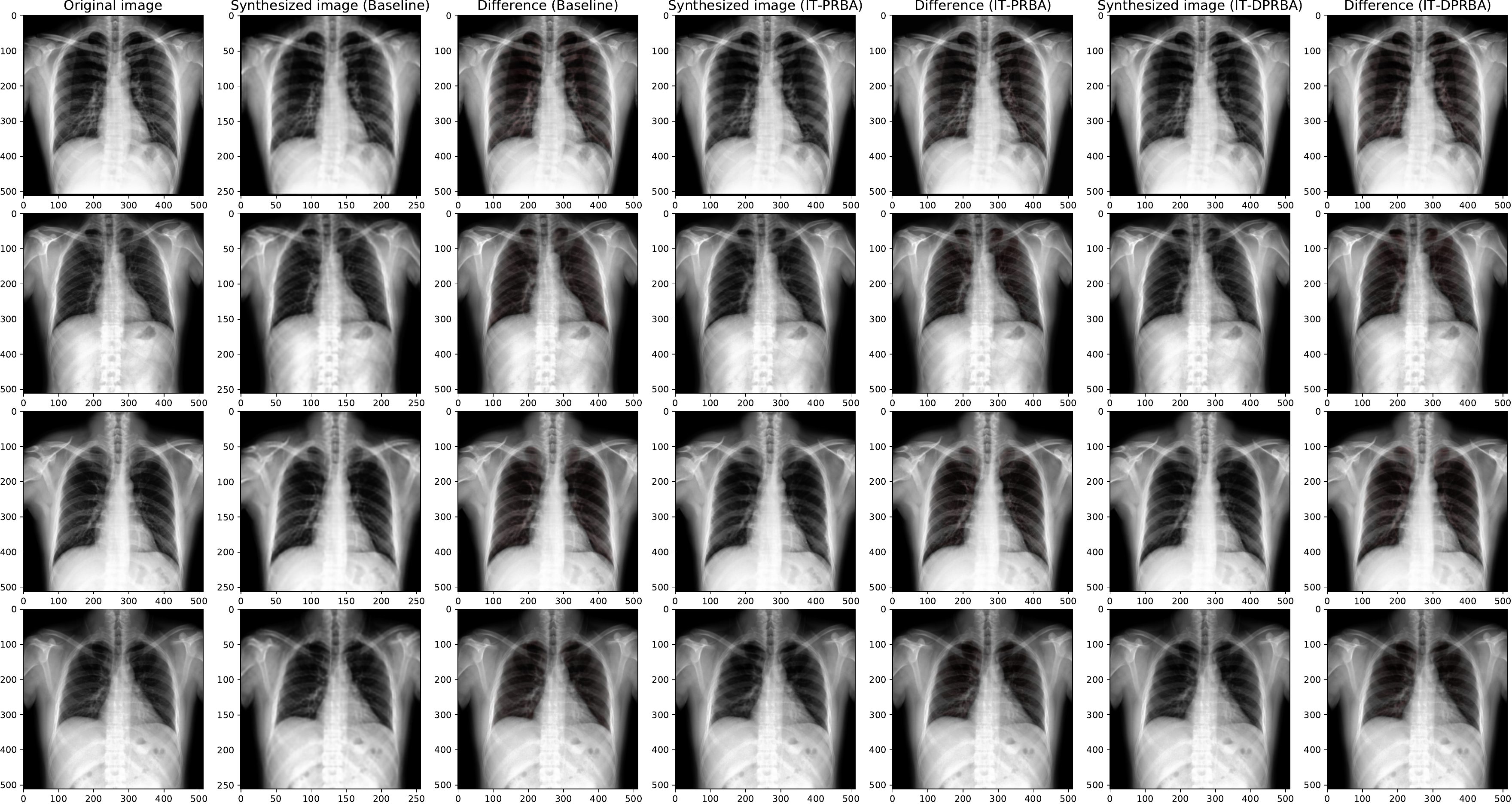}}
    \subfigure[]{\includegraphics[width=0.9\textwidth]{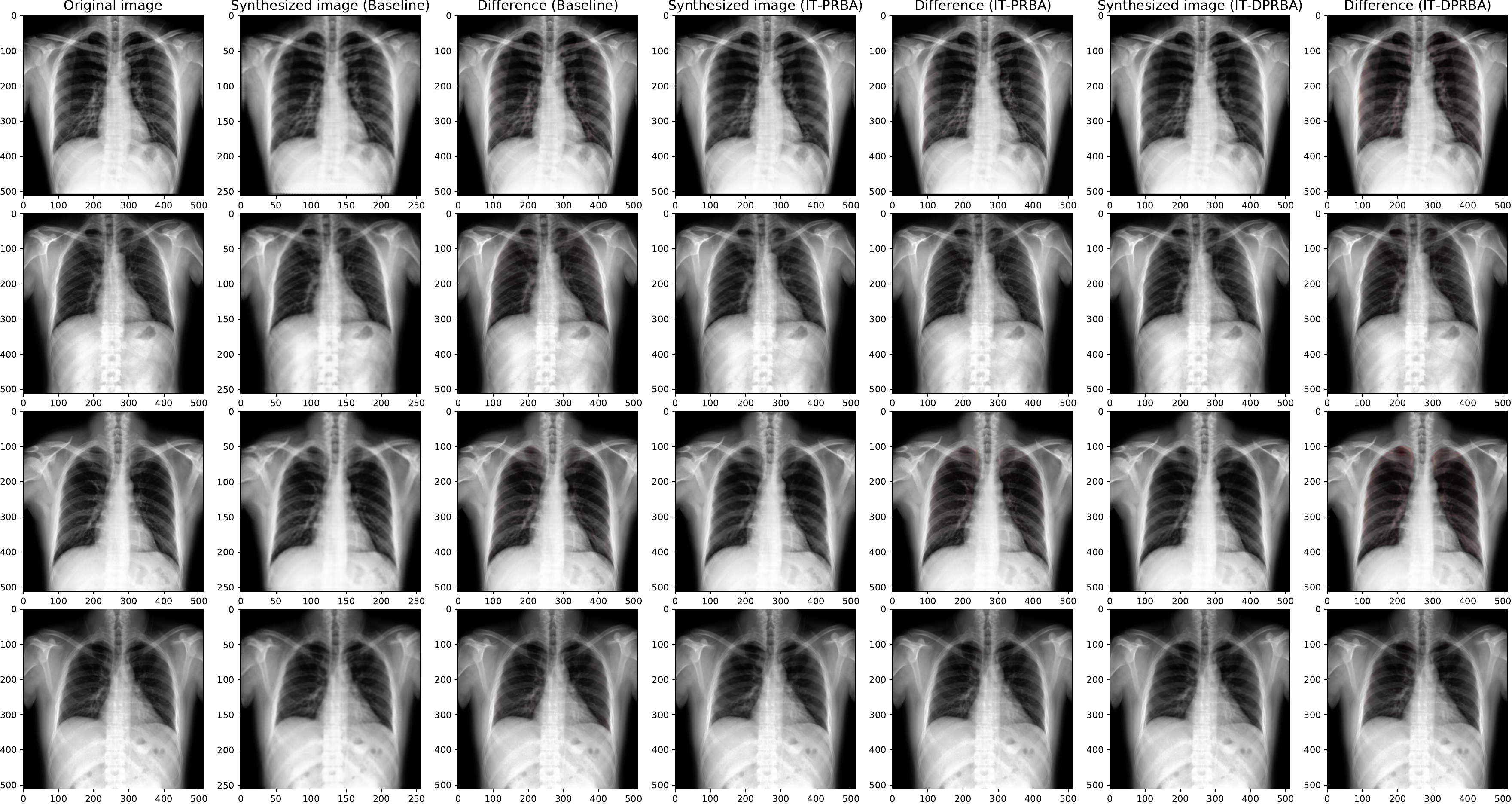}}
\caption{Comparison between anomaly localization maps $v_t$ obtained from proposed method and baseline for normal patients. All were commonly used based on (a) CUT or (b) CycleGAN.}
\label{fig_comp_cycle_normal}
\end{figure*}

\begin{figure*}[t]
\centering
	\includegraphics[width=0.95\textwidth]{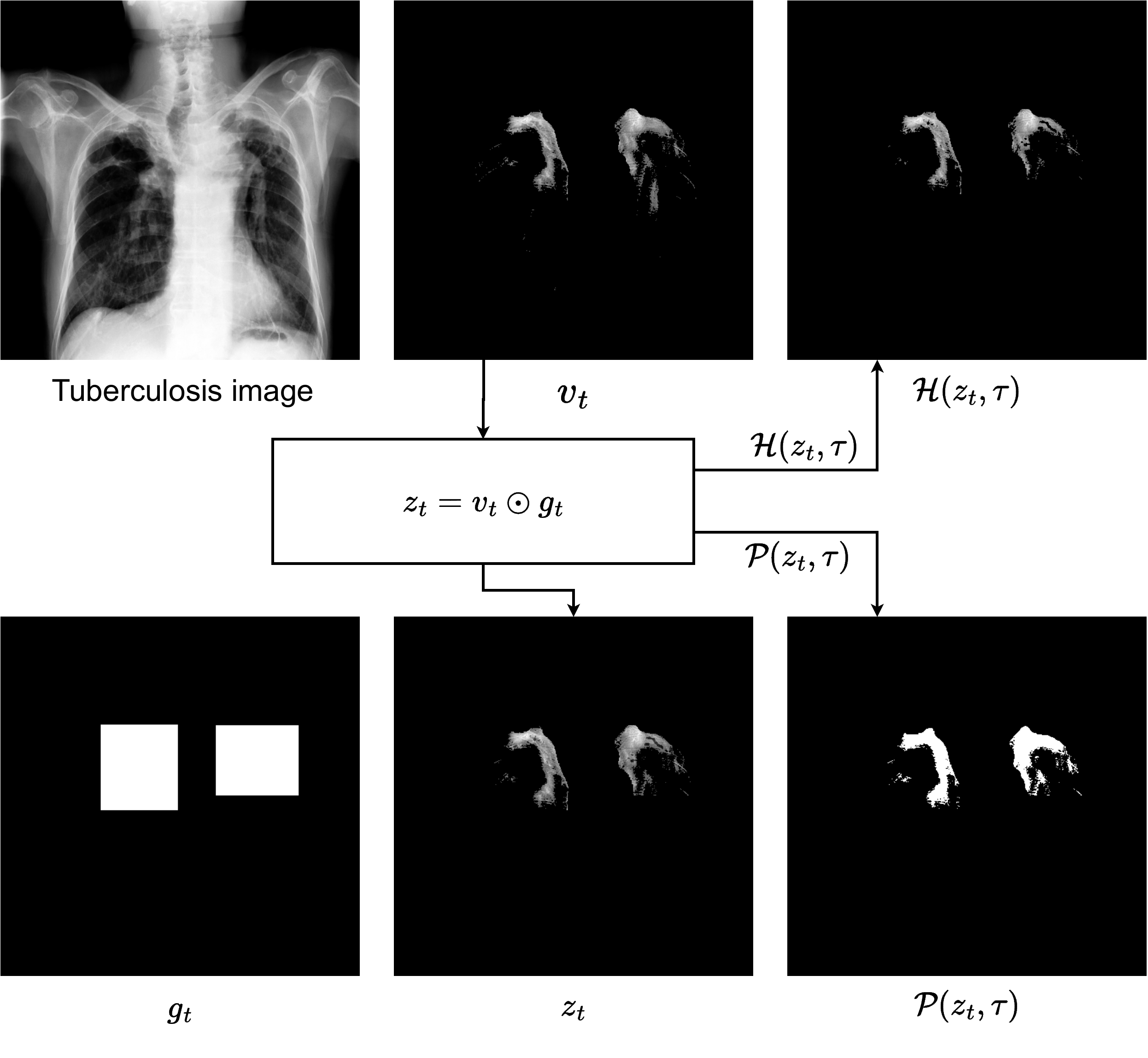}
	\caption{\footnotesize Illustration of how to derive metrics to compare and evaluate the performance of AI techniques for lung anomaly localization} 
	\label{fig:method_metric}
 \vspace{-0.2cm}
\end{figure*}

\begin{figure*}[h]
\centering
    \includegraphics[width=.8\textwidth]{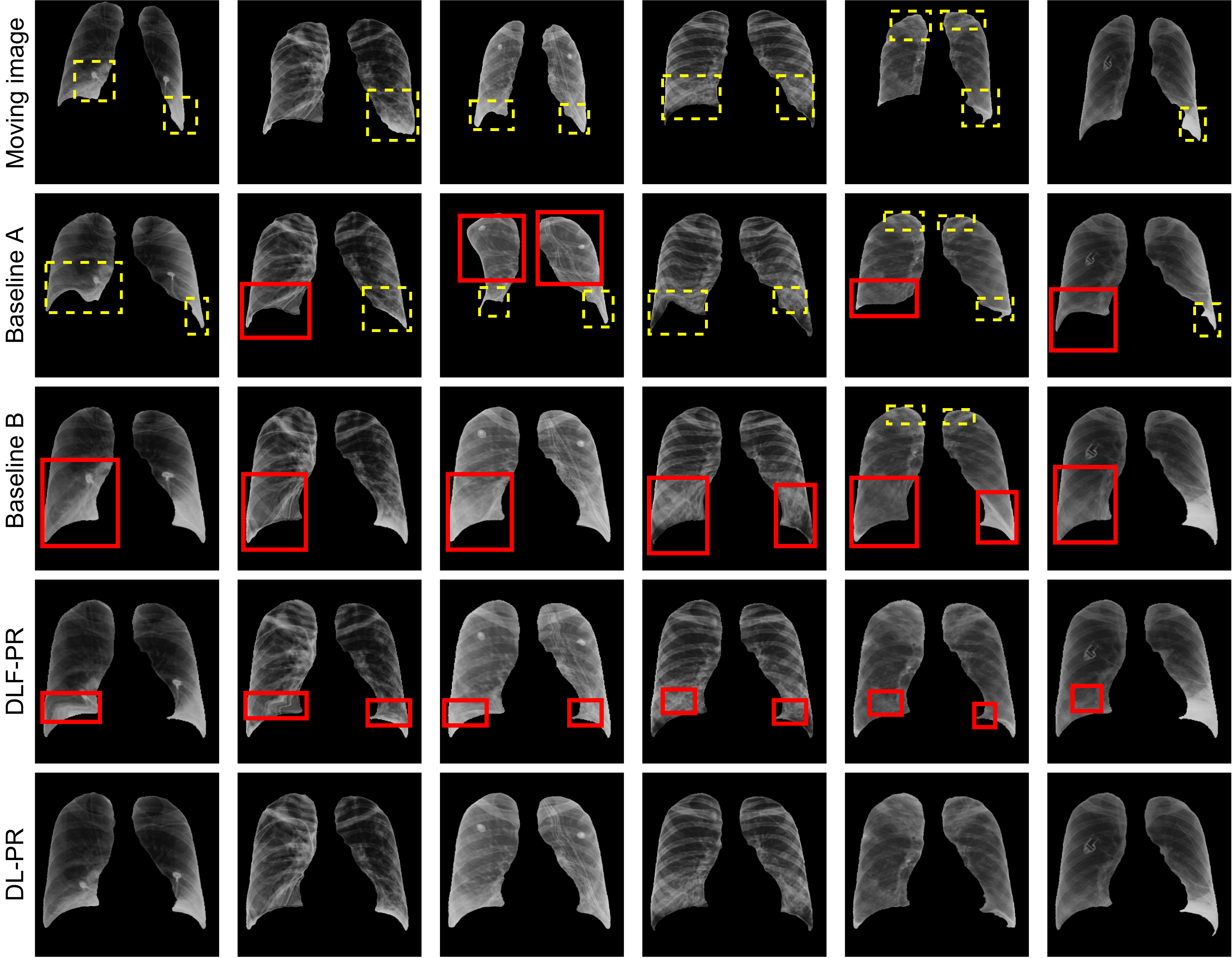}
\caption{Performance comparison between registration techniques for consolidation case. The registered/moved CXR images were obtained from the proposed DL-PR/DLF-PR and baselines A/B. Our registration approach has fewer or no artifacts.}
\label{registration_img_all_A}
\end{figure*}
 
\begin{figure*}[h]
\centering 
    \includegraphics[width=.9\textwidth]{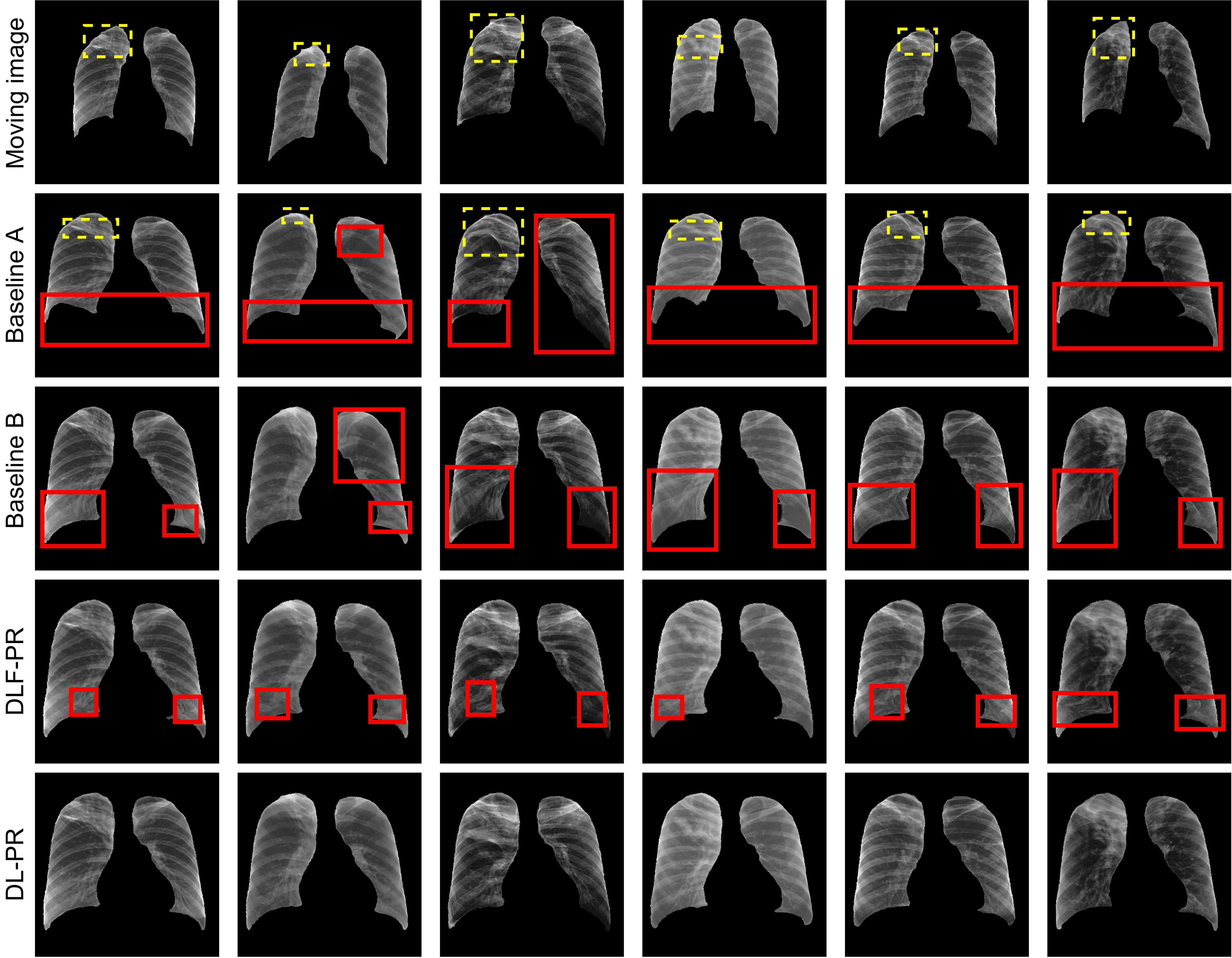}
\caption{Performance comparison between registration techniques for tuberculosis case. The registered/moved CXR images were obtained from the proposed DL-PR/DLF-PR and baselines A/B. Our registration approach has fewer or no artifacts.}
\label{registration_img_all_B}
\end{figure*}

\begin{figure*}[h]
\centering
    \includegraphics[width=.9\textwidth]{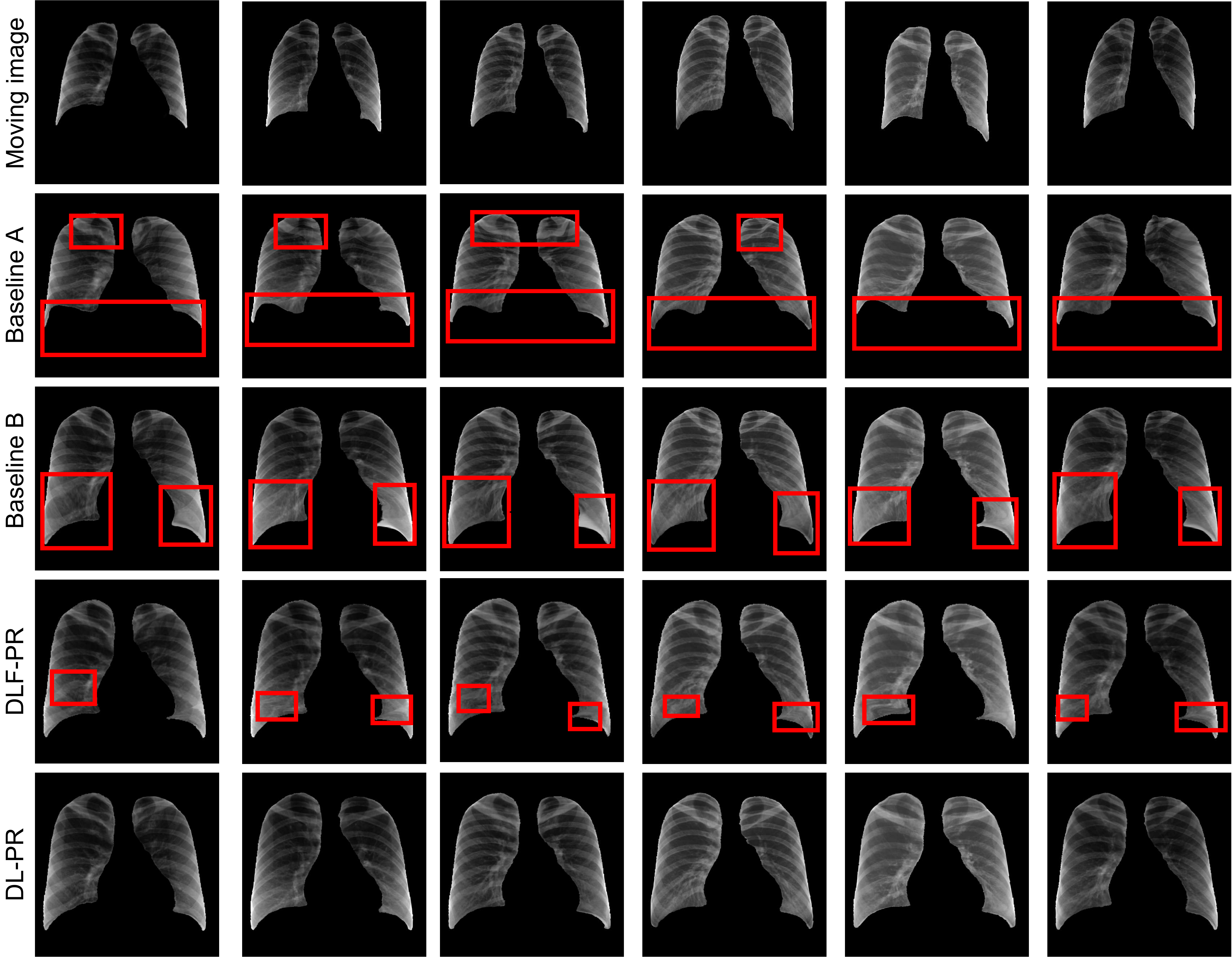}
\caption{Performance comparison between registration techniques for normal case. The registered/moved CXR images were obtained from the proposed DL-PR/DLF-PR and baselines A/B. Our registration approach has fewer or no artifacts.}
\label{registration_img_all}
\end{figure*}

\clearpage 
\bibliographystyle{unsrtnat}
\bibliography{ref.bib}

\end{document}